\documentclass[epj,onecolumn]{svjour}
\usepackage[intlimits]{amsmath}
\usepackage{amssymb}
\usepackage{graphicx}
\usepackage{color}
\usepackage{bbm}
\usepackage{booktabs}
\usepackage{mathrsfs,slashed}
\usepackage{hyperref}
\usepackage{epsfig}
\usepackage{tikz}
\usetikzlibrary{arrows.meta}

\newcommand{\bea}{\begin{eqnarray}}
\newcommand{\eea}{\end{eqnarray}}
\newcommand{\beq}{\begin{equation}}
\newcommand{\eeq}{\end{equation}}

\begin{document}

\title{
    Thermodynamics of quark matter with multiquark clusters in an effective Beth-Uhlenbeck type approach\thanks{This work has been initiated by Peter Schuck in discussions during the Workshop on "Light Clusters in Nuclei and  Nuclear Matter: Nuclear Structure and Decay, Heavy Ion Collisions and Astrophysics",  at the ECT* Trento, September 2 - 6, 2019.}
}
\titlerunning{Beth-Uhlenbeck approach to clusters in quark matter}
\author{
    D. Blaschke\inst{1,2,3}
    \and
    M.~Cierniak\inst{1}
    \and
    O.~Ivanytskyi\inst{4}
    \and
    G.~R\"opke\inst{1,5}
}
\authorrunning{D. Blaschke, M. Cierniak, O. Ivanytskyi and G. R\"opke
}
\institute{
    Institute of Theoretical Physics,
    University of Wroclaw,
    Max Born place 9,
    50-204 Wroclaw, Poland
    \and
    Helmholtz-Zentrum Dresden-Rossendorf (HZDR),
    Bautzner Landstrasse 400,
    01328 Dresden, Germany
    \and
    Center for Advanced Systems Understanding (CASUS),
    Untermarkt 20,
    02826 G\"orlitz, Germany
    \and
    Incubator of Scientific Excellence---Centre for Simulations of Superdense Fluids, University of Wroc?aw, 50-204, Wroclaw, Poland
    \and
Institute of Physics,
University of Rostock,
Albert-Einstein Str. 23-24,
18059 Rostock, Germany
}

\date{Received: \today / Revised version: date}

\abstract{
We describe multiquark clusters in quark matter within a Beth-Uhlenbeck approach in a background gluon field coupled to the underlying chiral quark dynamics using the Polyakov gauge. 
An effective potential for the traced Polyakov loop is used to establish the center symmetry of the SU(3) color which suppresses colored states and its dynamical breaking as an aspect of the confinement/deconfinement transition.
Quark confinement is modeled by a large quark mass in vacuum which is motivated by a confining density functional approach.
A multiquark cluster containing $n$ quarks and antiquarks is described as a binary composite of smaller subclusters $n_1$ and $n_2$ ($n_1+n_2=n$).
It has a spectrum consisting of a bound state and a scattering state continuum. 
For the corresponding cluster-cluster phase shifts we use simple ans\"atze that capture the Mott dissociation of clusters as a function of temperature and chemical potential. 
We go beyond the simple "step-up-step-down" model that ignores continuum correlations and introduce an improved model that includes them in a generic form.
In order to explain the model, we restrict ourselves here to the cases where the cluster size is $1 \le n \le 6$.
A striking result is the suppression of the abundance of colored multiquark clusters at low temperatures by the coupling to the Polyakov loop and their importance for a quantitative description of lattice QCD thermodynamics at non-vanishing baryochemical potentials. 
An important ingredient are Polyakov-loop generalized distribution functions of $n$-quark clusters which are derived here for the first time.
Within our approach we calculate thermodynamic properties such as baryon density and entropy. 
We demonstrate that the limits of a hadron resonance gas at low temperatures and $\mathcal{O}(g^2)$ perturbative QCD at high temperatures are correctly reproduced. 
A comparison with lattice calculations shows that our model is able to give a unified, systematic approach to describe properties of the quark-gluon-hadron system.
}
\date{\today}
\maketitle

\section{Introduction}
\label{sec:intro}

The phase diagram of quantum chromodynamics (QCD) is a subject of intense theoretical and experimental research.
Exploratory probes of the hot and increasingly dense nuclear matter are being performed in heavy--ion collision (HIC) experiments at the Relativistic Heavy-Ion Collider (RHIC) at BNL Brookhaven, the Large Hadron Collider (LHC) and the Super Proton Synchrotron (SPS) at CERN in Geneva with plans for many additional accelerator experiments in the near future.
In the low-temperature range of the phase diagram, the possibility of a phase transition to a cold and dense quark--gluon plasma (QGP) in neutron stars is widely considered (see \cite{Alford:2007xm,Anglani:2013gfu,Pisarski:2019cvo,Bauswein:2022vtq} for recent reviews).
Many ongoing observational efforts aim at detecting signs of a phase transition inside neutron star cores.
The current iteration of these multi-messenger observations include the NICER mission, and the LIGO--VIRGO collaboration with their gravitational wave detectors.
The efforts of the NICER team have produced vital data on the masses and radii of pulsars PSR J0030+0451 \cite{Riley:2019yda,Miller:2019cac} and PSR J0740+6620 \cite{Riley:2021pdl,Miller:2021qha} (currently the heaviest known neutron star with $2.08\pm 0.07~M_\odot$, as independent Shapiro--delay based mass measurements \cite{NANOGrav:2019jur}) while the analyses of the gravitational wave signal from the binary neutron star merger event GW170817 have put constraints on the tidal deformability of neutron stars in their typical mass range of $1.4~M_\odot$
\cite{LIGOScientific:2017vwq,LIGOScientific:2018cki,LIGOScientific:2018hze}.
Future measurements of postmerger gravitational waves may contribute to deciding the question whether a strong phase transition to deconfined quark matter may occur in those systems \cite{Bauswein:2018bma}.
In addition, the hypothesis of a phase transition induced shock revival in core--collapse supernova \cite{Fischer:2017zcr} might hint at an additional observational avenue of studying the structure of the hot and dense sector of the QCD phase diagram (cf. \cite{Bauswein:2022vtq} and references therein).

All of these efforts require a good theoretical understanding of the phase diagram of strongly interacting matter and its structure.
The only \textit{ab--initio} solutions of QCD, which could be applied to the study of the phase diagram are lattice-QCD (LQCD) calculations, where Monte--Carlo simulations of the QCD partition function are performed on a discretized Euclidean space--time lattice.
The vanishing baryochemical potential ($\mu_B=0$) sector of the QCD phase diagram is well explored within this approach, see, for instance, Ref. 
\cite{Borsanyi:2010bp,Borsanyi:2012cr,Bazavov:2014xya,Bazavov:2018mes}.
However, despite recent progress in calculating finite $\mu_B$ observables on the lattice \cite{Borsanyi:2021sxv,Bollweg:2022fqq,Bollweg:2022rps},
the available density range is still insufficient to provide theoretical data applicable to both HIC and neutron star studies due to the numerical sign problem for finite $\mu_B$ calculations.

Therefore, in order to explore the QCD phase diagram at arbitrary temperatures and densities one has to use effective models based on the underlying symmetries of QCD. 
One such model is the Nambu--Jona-Lasinio (NJL) model \cite{Nambu:1961tp,Nambu:1961fr}.
It is extensively used in studies of dynamical chiral symmetry breaking and restoration (cf. \cite{Buballa:2003qv}). 
However, this model does not capture the aspect of quark confinement and leads to an unphysical dominance of colored quark degrees of freedom already at relatively low temperatures, below 100 MeV. 
Therefore, an extension of the NJL model by an additional coupling between the fermionic degrees of freedom and a gluon background field in the Polyakov gauge (or PNJL model \cite{Fukushima:2003fw,Ratti:2005jh,Roessner:2006xn,Roessner:2007gha}) has been introduced which addresses the aspect of quark confinement by a statistical suppression of the contribution of free quarks to thermodynamic observables at low temperatures and chemical potentials.
In this work we further extend the PNJL--based QGP model by a perturbative QCD correction of $\mathcal{O}(\alpha_s)$ 
in the expansion w.r.t. the strong fine structure constant $\alpha_s=g^2/(4\pi)$ stemming from one-gluon exchange which, following Refs. 
\cite{Blaschke:2020lhm} (for details, see also \cite{Blaschke:2016hzu,Blaschke:2016fdh})
applies only to processes with high momentum transfer $p>\Lambda_{\rm pert}$.
This extension of the PNJL model leads to an improvement of the comparison with LQCD thermodynamics at high temperatures.

The model is defined to allow for a thermodynamic description of matter at finite temperature $T$ and baryochemical potential $\mu_B$ 
under conditions of isospin symmetry and strangeness neutrality.

For a realistic description of strong interacting matter, the effective model should take into account quark hadronization in the domain of low $T$ and $\mu_B$, where the resulting hadrons are composite quark clusters. 
In the spirit of \cite{Blaschke:2020lhm}, the composite nature of the low density hadronic degrees of freedom is accounted for by employing a generalized Beth-Uhlenbeck formula \cite{Blaschke:2016hzu,Blaschke:2016fdh,Hufner:1994ma,Zhuang:1994dw,Wergieluk:2012gd,Yamazaki:2012ux,Blaschke:2013zaa,Yamazaki:2013yua,Torres-Rincon:2016ahl,Torres-Rincon:2017zbr,Blaschke:2016sqn}, which provides an interface between the low density and low temperature hadronic matter and the high density and high temperature PNJL-based QGP.
This approach to quark clustering closely resembles the idea of the hadron--resonance--gas (HRG) model, which is crucial to the interpretation of LQCD calculations at low temperatures.

The HRG model assumes that the strong interactions are saturated by hadron resonance formation  \cite{Hagedorn:1965st}.
Indeed, this assumption produces a description of low energy QCD matter as a free HRG well in agreement with both LQCD studies below the pseudocritical temperature $T_{c}$ \cite{Borsanyi:2011sw,HotQCD:2012fhj,Bellwied:2015lba,Bellwied:2013cta} and HIC experiments (cf. \cite{Andronic:2017pug} and reference therein).

Unsurprisingly, however, at increasing temperatures this agreement falters due to the rising importance of the quark substructure of hadrons leading to repulsive interactions, which the unmodified HRG model does not take into account. 
One fact is that highly excited resonances in the continuum of scattering states cannot be treated like bound states. These weak correlations are seen in the phase shifts and give only a small contribution to the virial expansion of the thermodynamic functions, as known, for instance from the Beth-Uhlenbeck formula discussed below.

Another fact is that the model of ideal mixture of non-interacting clusters fails when density is increasing.
Various methods of solving this problem exist, for instance by a hadronic excluded volume. 
One physical mechanism for the repulsive interactions in a hadron gas is seen in the quark substructure of hadrons which, following the Pauli exclusion principle, leads to a repulsive interaction by quark Pauli blocking among hadrons \cite{Ropke:1986qs,Blaschke:2020qrs}. 
These effects of exchange symmetry are precursory to the delocalization of the hadron wave functions in the hot, dense hadron gas which entails the dissociation of hadrons in the analogue of a  Mott effect introduced in nuclear physics \cite{ROPKE1982536} and known from solid state and plasma physics \cite{Kraeft:1986}.
In order to provide a description of the QGP on the same microscopic footing with the gas of hadronic resonances as bound states of a confining interquark potential, the principle of saturation of the color charge within a range of nearest neighbors has been applied within the string-flip model (SFM) \cite{Horowitz:1985tx,Ropke:1986qs}. 
In Ref. \cite{Yukalov:1997jk}, an improvement of the SFM has been presented which includes relativistic kinematics for the quarks and discusses a mixed phase between the SFM plasma and a schematic HRG model containing multi-quark clusters. The results were compared with the early lattice QCD simulations of that time. 
However, a consistent formulation of medium effects on bound state formation is absent in this approach so that this mixed phase construction can only be viewed as an early realization of the switch function concept of Ref. \cite{Albright:2014gva}.
The SFM model for the QGP has been considerably improved in the subsequent work of Khvorostukhin et al. \cite{Khvorostukin:2006aw}, where a confining quasiparticle model of quarks and gluons has been introduced in the form of a relativistic density functional (RDF) approach. 
The form of this density functional has later been generalized to include the coupling to a nonlinear vector mean field that allowed to discuss heavy hybrid neutron stars and mass twin solutions for them \cite{Kaltenborn:2017hus}. The most recent development of the SFM is a lagrangian RDF formulation that obeys chiral symmetry and contains vector meson and diquark mean fields and is shown to obey modern constraints from multi-messenger observations of neutron stars and their mergers \cite{Ivanytskyi:2022oxv}.

However, to provide qualitative insight into the transition from an HRG to a QGP, in the present work we abstain from 
a detailed microscopic model for solving the few-quark problem in a dense medium.
For our study, we will incorporate the effect of the repulsive interaction as well as the eventual temperature and density dissociation of hadrons (the Mott effect) directly into a model for the phase shift $\delta_i$ of a hadron species $i$ in the generalized Beth-Uhlenbeck formula.
The ansatz will be similar to the one used in \cite{Blaschke:2016hzu,Blaschke:2016fdh}, in agreement with the Levinson theorem.
The resulting contribution to the thermodynamic observables will be referred to as the Mott--hadron--resonance--gas (MHRG) contribution. The full MHRG+PNJL model embodies the main consequence of quark chiral symmetry restoration, i.e. the lowering of the continuum thresholds for multiquark scattering states, which entails the dissociation of multiquark bound states
(Mott effect).
A general approach to strong correlations in many-particle systems is the $\Phi$-derivable approach which is based on the Luttinger-Ward-Baym functional \cite{Luttinger:1960ua,Baym:1961zz,Baym:1962sx}
and has recently been applied to a $T$-matrix description of the QGP thermodynamics as well as spectral and transport properties by Liu and Rapp \cite{Liu:2017qah}. However, these authors did not extend the approach to include bound state formation and the limiting case of the hadron resonance gas phase. This is the main goal of the present work, where 
we generalize the $\Phi$-derivable approach to deal with multi-quark clusters.

As an extension of the previous work \cite{Blaschke:2020lhm}, we have considered clusters beyond the color--singlet states, and we allow for any combination of quarks and antiquarks up to 6 valence degrees of freedom.
In this way, we also take into account all colored multi-quark cluster states which, however, are strongly suppressed in the region below the Mott temperature due to their coupling to the Polyakov loop.

As a benchmark for defining the model, we use continu-\-um--extrapolated LQCD calculations of the
baryon density and the entropy density as a function of the temperature for finite values of 
$\mu_B/T$ \cite{Borsanyi:2021sxv,Bollweg:2022fqq}.
These data allow us to test the PNJL model with the 
MHRG virial correction terms.
We perform exploratory calculations accounting 
for a limited number of resonances from the Particle Data Group (PDG) up to a mass threshold of about $2600$ MeV.

Furthermore, instead of deriving the quark mass self--consistently, we make the assumption of a sudden drop of the quark mass at the $\mu_B$-dependent pseudocritical temperature $T_c(\mu_B)$ from the constituent quark mass value to the current quark mass value.
This assumption entails that the pseudocritical temperature can be identified with the Mott temperature 
$T_{\rm Mott}(\mu_B)$ where all hadronic bound states dissociate into their quark constituents.
This allows us to focus on the cluster contribution to the Polyakov--loop calculation, without taking into account the complicated interplay between the singlet clusters, the scalar mean field and the Polyakov loop.
Such an analysis is definitely warranted, especially due to the discrepancy between the pseudocritical temperature of the chiral crossover transition obtained in LQCD simulations and Polyakov--loop generalized chiral quark models. 
But this issue is beyond the scope of the present work.

This work is organized as follows: In Sect. \ref{sec:GBU} we present the theoretical aspects of our approach, i.e., the QGP thermodynamics based on the PNJL thermodynamic potential, and the treatment of clusters in the context of the MHRG approach. A major issue is the determination of the scattering phase shifts, which follow in principle from the solution of a few-quark in-medium wave equation, but here are postulated in a generic, medium-dependent form in Sect. \ref{sec:model} along with other parameters of the model. Results for the thermodynamic quantities baryon density and entropy as a function of $T$ and $\mu_B$ are presented and discussed in Sect. \ref{sec:results}. 
In Sect. \ref{sec:summary}, a summary is given and conclusions are drawn. Technical details are collected in Appendices A and B.

\section{Generalized Beth-Uhlenbeck approach to quark clustering in the Polyakov gauge}
\label{sec:GBU}

In order to satisfactory reproduce the high temperature lattice QCD predictions, we shift now towards an effective Polyakov-loop quark--gluon plasma (QGP) model based on the quark quasiparticle description 
combined with the Polyakov--loop potential $\mathcal{U}(\phi,\bar{\phi})$ and a $\mathcal{O}(\alpha_s)$ perturbative virial correction. 
In an effort to maintain the description of hadrons as nonperturbative correlations of quarks and gluons, we couple this model to the Mott--hadron resonance gas (MHRG), in analogy to \cite{Blaschke:2020lhm}, and extended it to finite chemical potentials.

In this work we develop the technique of the cluster virial expansion approach for multi-quark clusters at finite temperatures and chemical potentials on the basis of a cluster generalization of the so-called $\Phi$-derivable approach. 
Within a Green's function approach \cite{Baym:1961zz,Baym:1962sx} the correlation functional $\Phi$ is introduced where the contributions of interaction are represented by closed-loop diagrams.
When for the $\Phi$ functional a restriction to closed two-loop diagrams in cluster Green's functions is applied, this approach is equivalent to the generalized Beth-Uhlenbeck approach to clustering in hot, dense Fermi-systems
\cite{Ropke:2012qv}.
The intricacy of color confinement in low-density quark matter is considered by coupling the quarks and their clusters to the Polyakov-loop background field which serves to suppress the appearance of colored clusters in the region of quark confinement.
We demonstrate a satisfactory comparison of our results for the thermodynamics of clustered quark matter with recent lattice QCD simulations at finite temperature and chemical potentials, where they are available.

We introduce an ansatz for the thermodynamic potential which separates the QGP sector of quark and gluon quasiparticles from that of the MHRG comprised of hadrons which are understood as quark bound states (multiquark clusters) that can undergo a Mott dissociation,
\begin{equation} 
\label{eq:omega_total}
    \Omega(T,\mu,\phi,\bar{\phi})=\Omega_{\rm QGP}(T,\mu,\phi,\bar{\phi})+\Omega_{\rm MHRG}(T,\mu,\phi,\bar{\phi}).
\end{equation}
In the following subsections we explain these two contributions more in detail. 

\subsection{QGP thermodynamics of quarks and gluons}

The thermodynamics of the deconfined QGP can be separated into a perturbative and a nonperturbative contribution
\begin{equation}
\Omega_{\rm QGP}(T,\mu,\phi,\bar{\phi})=\Omega_{\rm NP}(T,\mu,\phi,\bar{\phi})+\Omega_{\rm pert}(T,\mu,\phi,\bar{\phi}), 
\end{equation}
where $\Omega_{\rm NP}$ describes the nonperturbative low-energy QCD quark and gluon degrees of freedom via the mean--field thermodynamic potential 
\begin{equation}
\label{eq:OmegaNP}
\Omega_{\rm NP}(T,\mu,\phi,\bar{\phi})=\Omega_{Q}(T,\mu,\phi,\bar{\phi})+\mathcal{U}(\phi,\bar\phi) 
\end{equation}
and perturbative corrections are absorbed in  
$\Omega_{\rm pert}(T,\mu,\phi,\bar{\phi})$.

We take into account the minimal coupling of quarks to a homogeneous gluon background mean field  in the Polyakov gauge $A_4=\lambda_3 A_4^3 + \sqrt{3} \lambda_8 A_4^8$ 
while quark interactions are accounted for by a suitably chosen relativistic energy density functional $\mathcal{L}_{\rm int}$
\begin{eqnarray}
\label{eq:Dirac}
 \mathcal{L}_0 &=& \bar{q} \left[\gamma_0(i\omega_n + \mu -iA_4) -\vec{\gamma} \cdot \vec{p} - m_0 \right] q 
 \nonumber\\
&&-  \mathcal{U}(\phi,\bar\phi) - 
\mathcal{L}_{\rm int}~,
\end{eqnarray}
with 
$\mathcal{U}(\phi,\bar\phi)$ denoting the Polyakov--loop (gluon) potential for which we will not make use of the traditional form found in \cite{Ratti:2005jh}, but the one given in \cite{Lo:2013hla} as explained below.
The traced Polyakov loop is defined as 
\begin{equation}
    \phi = {\rm Tr}_c \exp(i\beta A_4)/N_c,
\end{equation}
where $N_c=3$ is the number of colors in the $SU(N_c)$ gauge theory and $\beta=1/T$ is the inverse temperature.
Abbreviating this expression in accordance with the standard notation found in the literature, we have $A_4^3=\varphi_3$ and $A_4^8=\varphi_8$, so that $A_4={\rm diag} (\varphi_3+\varphi_8,-\varphi_3+\varphi_8,-2\varphi_8)$ and finally
\begin{equation}
\label{eq:PL}
\phi = \frac{1}{N_c}\left({\rm e}^{-2i\beta\varphi_8}+{\rm e}^{i\beta(\varphi_8-\varphi_3)}+{\rm e}^{i\beta(\varphi_8+\varphi_3)} \right)~,~
\bar{\phi} = \phi^*~.
\end{equation}   

The minimal coupling of the quark to the homogeneous gluon background field in the Polyakov gauge leads to a shift of the Matsubara frequency 
$i\omega_n \to i\omega_n - iA_4$.
We define the distribution function of a quark of given color $c$ as the Matsubara sum over the quark propagator
\begin{equation}
F_c (E_p - \mu + i A_4) \equiv \frac{1}{\beta} \sum_n \frac{1}{ (i\omega_n - E_p + \mu)\mathbbm{1}_c - i(A_4)_{cc}} ~.
\end{equation}
The Polyakov-loop distribution function is then obtained after performing the color trace
\begin{eqnarray}
f^{(1),+}_\phi &=& {\rm Tr}_c F_c (E_p - \mu + i A_4)
\\
&=& \sum_{c=1}^{3} f[(E_p - \mu) + i(A_4)_{cc}] ,
\end{eqnarray}
where $f[(E_p - \mu)]$ is the ordinary Fermi function which can be written as
\begin{equation}
f(E_p - \mu) \equiv -\frac{1}{\beta} \frac{\partial z^+}{\partial E_p} , 
\end{equation}
with $z^+ \equiv \ln [1 + y^+_1 ]$ and
$y^+_1={\rm e}^{-\beta (E_p - \mu)}$.
Now we can write 
\begin{eqnarray}
f^{(1),+}_\phi &=&  \frac{1}{\beta} \sum_{c=1}^3 
 \frac{\partial \ln \left[1 + y^+_1{\rm e}^{-i\beta (A_4)_{cc}}\right]}{\partial E_p}\\
 &=& \frac{1}{\beta} \frac{\partial}{\partial E_p}  \ln \left[1 + 3 (\bar{\phi} + \phi y^+_1) y^+_1 + {y^+_1}^3\right]
 \end{eqnarray}
and obtain the Polyakov-loop generalized single-quark distribution function 
 \begin{eqnarray}
 \label{eq:f-phi}
f^{(1),+}_\phi &=& \frac{(\bar{\phi} + 2\phi y^+_1) y^+_1 + {y^+_1}^3}{1 + 3 (\bar{\phi} + \phi y^+_1) y^+_1 + {y^+_1}^3}~.
\end{eqnarray}
We note that the distribution function for an antiquark $\left[f^{(1),-}_\phi\right]^*$ is obtained from Eq.~(\ref{eq:f-phi}) by the replacements: $\mu \to - \mu$ and $\phi \to \bar\phi=\phi^*$.
The nonperturbative part of the QGP thermodynamics (\ref{eq:OmegaNP}) consists of a quark quasiparticle contribution and the Polyakov--loop potential $\mathcal{U}$ 
for which we adopt the form defined in \cite{Lo:2013hla},
\begin{eqnarray} \label{eq:Omega_gluon}
    \mathcal{U}(T,\phi,\bar{\phi})
    &=&
    T^4
    \left[
    -\frac{1}{2}a(T)\bar{\phi}\phi+b(T)\log M_H \right. \\
    &&\left. +\frac{1}{2}c(T)\left(\phi^3+\bar{\phi}^3\right)+d(T)\left(\bar{\phi}\phi\right)^2
    \right] \nonumber
\end{eqnarray} 
with $M_H$ being the SU(3) Haar measure
\begin{equation}
    M_H=1-6\bar{\phi}\phi+4\left(\phi^3+\bar{\phi}^3\right)-3\left(\bar{\phi}\phi\right)^2,
\end{equation}
and
\begin{equation}    a(T)=\frac{a_1+a_2\left(T_0/T\right)+a_3\left(T_0/T\right)^2}{1+a_4\left(T_0/T\right)+a_5\left(T_0/T\right)^2}
+a_6\frac{\mu^2 T_0^2}{T^4},
\end{equation}
\begin{equation}
    b(T)=b_1\left(T_0/T\right)^{b_4}\left[1-e^{b_2\left(T_0/T\right)^{b_3}}\right],
\end{equation}
\begin{equation}    c(T)=\frac{c_1+c_2\left(T_0/T\right)+c_3\left(T_0/T\right)^2}{1+c_4\left(T_0/T\right)+c_5\left(T_0/T\right)^2},
\end{equation}
\begin{equation}
d(T)=\frac{d_1+d_2\left(T_0/T\right)+d_3\left(T_0/T\right)^2}{1+d_4\left(T_0/T\right)+d_5\left(T_0/T\right)^2}.
\end{equation}
The parameters $a_1\dots a_5$, $b_1\dots b_4$, $c_1\dots c_5$, $d_1\dots d_5$ are taken from the pure SU(3) lattice gauge fit performed in \cite{Lo:2013hla}. 
The parameter $a_6=1.28$ is 
introduced here to improve the description of the
baryon density in comparison with LQCD results \cite{Borsanyi:2021sxv}.
The parameter $T_0$ is related to the critical temperature for deconfinement in the case of pure gluodynamics, where the value $T_0=270$ MeV is taken from pure gauge theory simulations on the lattice. One has to invoke a flavor dependence of this parameter \cite{Schaefer:2007pw}, which for our applications to the realistic case of $N_f=2+1$ flavors is chosen to be $T_0=205$ MeV in order to provide the best agreement with the LQCD data \cite{Borsanyi:2021sxv} discussed below.

The mean field thermodynamics in the quark sector of the PNJL model is described by
\begin{eqnarray} 
\label{eq:Omega_quark}
    \Omega_{Q}(T,\mu,\phi,\bar{\phi})
    &=&
    \sum_{f=u,d,s}\frac{\Delta^2_f(T,\mu)}{8G_s}
    -2N_fN_c
    \int_0^\Lambda\frac{dp\, p^2}{2\pi^2} E_p 
    \nonumber\\
    &-&N_fN_c
    \int\frac{dp\, p^2}{3\pi^2}
    \frac{p}{E_p}    \left\{
f^{(1),+}_\phi +\left[f^{(1),-}_\phi\right]^*
    \right\}.
    \nonumber\\
\end{eqnarray} 
Here, $G_s$ is the scalar coupling constant, and
$\Delta_f(T,\mu)=M_f(T,\mu)-m_f$ is the dynamically generated quark mass gap for the flavor $f$, see Tab. \ref{tab:0}.

\begin{table}[thb]
\centering
\begin{tabular}{|c|c|c|c|c|}
\hline
$B_f$ & quark & $M_f$ [MeV]& $m_f$ [MeV] & $d_f=2N_cN_f$ \\
\hline
1/3 & q=u,d &$627$ &5.6& $12$ \\
1/3 & s &$770$  &124& $6$ \\
\hline
\end{tabular}
\caption{Properties of quarks: $B_f$ - baryon number, $d_f$ - degeneracy, $M_f$ - constituent mass and $m_f$ - current mass of quark flavor $f$.
}
\label{tab:0}
\end{table}

The same structure for the mean field thermodynamic potential (\ref{eq:Omega_quark}) is also obtained in the confining relativistic density functional (RDF) approach, see Eqs. (16) and (17) of \cite{Ivanytskyi:2022oxv}, since it can be mapped to an NJL model with medium dependent couplings.  
Solutions of the gap equation for the medium dependent quark mass in this case are shown, e.g., in  \cite{Ivanytskyi:2021dgq} and exhibit a significantly larger constituent quark mass than NJL models, realising the effective quark confinement in the RDF approach.
In this work, we will use a sudden switch model
that entails a corresponding behavior of the thresholds for multi-quark continuum states as discussed in more detail in subsection \ref{ssec:quarkmass} below.

The divergence of the integral within $\Omega_Q$ is a known deficiency of the (P)NJL model. We regularize the divergent part of this integral by introducing a 3-dimensional momentum cutoff.
The cutoff $\Lambda$ is chosen such, that
\begin{equation}
    \frac{N_fN_c}{\pi^2}
    \int^{\Lambda(T,\mu)}_0
    dp\, p^2 \, E_p
    -\sum_{f=u,d,s}\frac{\Delta^2_f(T,\mu)}{8G_s} =  0,
\end{equation}
simplifying the quasiparticle expression for the quark thermodynamical potential to
\begin{equation}
    \Omega_{Q,{\rm reg}}
    =
    -N_fN_c
    \int\frac{dp\, p^2}{3\pi^2}
    \frac{p}{E_p}
    \left\{
    f^{(1),+}_\phi
    +\left[f^{(1),-}_\phi\right]^*
    \right\}.
\end{equation}
In addition, we introduce a perturbative virial correction contribution in two loop order, $\Omega_{\rm pert}(T,\mu,\phi,\bar{\phi})$ with a standard expression (cf. \cite{Kapusta:1989tk}, see also \cite{Turko:2011gw}) of the form
\begin{eqnarray}
\label{eq:Omega_pert}
    \Omega_{\rm pert}(T,\mu,\phi,\bar{\phi})=&
    \frac{8}{\pi}
    \frac{\alpha_s N_f}{\beta^4}&
    \left\{
    \frac{1}{6}\left(I^++I^-\right)
    \right.\\&&
    \left.
    +\frac{1}{4\pi^2}
    \left(I^++I^-\right)^2
    \right\} \nonumber
\end{eqnarray}
where
\begin{equation}
    I^\pm(T,\mu,\phi,\bar{\phi})
    =
    \int^\infty_{\Lambda_{\rm pert}/T}dx\frac{x}{3}f^{(1),\pm}_\phi
\end{equation}
are modified integrals introduced in \cite{Blaschke:2020lhm} and extended to finite chemical potentials with the generalized Polyakov--loop generalized Fermi distribution functions.
The infrared cutoff, $\Lambda_{\rm pert}=222$ MeV, represents the momentum range below which nonperturbative physics dominates.
This value is adjusted in order to provide agreement with the data on the LQCD thermodynamics (see Section \ref{sec:results}).
We use a temperature and chemical potential dependent regularized running coupling
\cite{Blaschke:2005jg,Shirkov:2002td,Peshier:2003ah}
\begin{eqnarray}
    \alpha_s(T,\mu) &=&
    \frac{12\pi}{11N_c-2N_f}
    \left[
    \frac{1}{\ln\left(x^2\right)}
    -\frac{1}{x^2-1}
    \right],
    \\
    x^2&=&
    \left(\frac{T}{T_*}\right)^2+\left(\frac{\mu}{\mu_*}\right)^2,
\end{eqnarray}
with $N_c=3$, $N_f=3$, $T_*=93.75$ MeV.
We choose $\mu_*=3\pi T_*$ in accordance with the argument of the switch function in \cite{Albright:2014gva}.

The traced Polyakov loop variables $\phi$ and $\bar{\phi}$
are obtained by solving the corresponding gap equations derived by minimizing 
the total thermodynamic potential. 
This includes the PNJL and perturbative parts of the QGP model, which were derived in \cite{Blaschke:2020lhm} for $\mu=0$. 
As an extension of that work, we have introduced finite--$\mu$ descriptions of each of the contributions to the thermodynamic potential.
Additionally, in the MHRG part we have permitted the contributions of color--triplet and color--antitriplet clusters, which, 
while strongly suppressed in the HRG phase where the traced Polyakov--loop is close to zero, have an impact on the Polyakov--loop behaviour and on the thermodynamics in the phase where the approximate chiral symmetry is restored and bound states underwent Mott dissociation. 
The results of these calculations will be shown in the following subsection.

\subsection{Clusters in quark matter and the MHRG approach
}

Below the deconfinement transition, quark matter appears predominantly as color-neutral clusters, the hadrons. 
The quark substructure of the hadronic system determines the interaction between hadrons, but also intrinsic excitations. 
The inclusion of cluster formation, including excited states and resonances is an indispensable ingredient to the thermodynamics of quark-gluon matter.

In contrast to plasma physics where we start from a known hamiltonian and use for the evaluation of the partition function well-elaborated methods such as partial summation of ladder diagrams to introduce bound state formation, 
this first principle approach is presently not in reach in QCD. 
Data from Lattice gauge theory simulations are not available for nuclear matter in the hadronic phase. 
Usually, this problem is circumvented by interpolation between two different approaches to describe high energy density matter,
the QGP at high temperatures as described above, and the hadron-resonance gas (HRG) below the deconfinement transition.
For instance, in \cite{Albright:2014gva} both approaches are matched together, using a switching function for a smooth transition. 
There, the crossover region at $\mu = 0$ occurs around $T = 170$ MeV and goes to $T=0$ around $\mu = 1.25$ GeV. 
In addition, the HRG is modified by an excluded-volume ansatz to mimic the interaction between hadrons in the hadronic phase.
Adapting parameters for this semi-empirical approach, in Ref. \cite{Albright:2014gva} good agreement with lattice data is obtained.
The introduction of a switching function "by hand", however, does not provide any insights to the microphysical processes that underlie the hadron-to-QGP transition.

We intend to find a unified description for the 
quark-gluon-hadron matter system including cluster formation and dissociation. 
The hadrons are obtained as bound states, but their disappearance at high densities is described as melting owing to medium effects. Instead of a phenomenological excluded volume concept, a quark exchange process may be considered as origin of repulsion. 
Thus, a main point is the account for antisymmetrization of the fermionic quark states what leads to Pauli blocking and the dissolution of bound states, denoted above as Mott effect.
In the absence of a fundamental hamiltonian of the system, we will use multiquark correlation functions to encode and model the dynamical properties of the system.

In this work we are interested in the systematic description of correlations in the quark system which are seen from the bound state formation, but also in the continuum correlations. 
For the two-particle system, continuum correlations are related to the scattering phase shifts as known from the Beth-Uhlenbeck formula. 
A generalization of this Beth-Uhlenbeck formula has been discussed \cite{Ropke:2012qv} where the scattering of two clusters was considered. 

When considering more than two elementary constitu\-ents of a cluster, different channels are possible where this cluster appears. 
Here, we deal with binary processes for the formation and the decay of the clusters into two subclusters. 
We consider two-component scattering phase shifts which indicate the occurrence of a resonance, but reflect also the possible existence of a bound state of these constituents.

In general, we have different possibilities for the decomposition of a cluster into subclusters. 
This is of relevance for the decays into various channels with corresponding branching ratios, 
but not for the thermodynamic properties which are determined by the density of states. For the thermodynamic properties it is sufficient to consider only a specific binary decomposition.
This approach is a generalization of the two-particle problem to a many-particle approach and has also been applied in nuclear physics to describe cluster formation and resonances in nuclear matter \cite{Ropke:2014fia,Ropke:2020peo}.

Within our approach, we consider all possible clusters formed from light quarks of flavors $f=u,d,s$, including also colored clusters. 
This is necessary to describe the transition to the QGP state. The set of clusters considered in the present work is shown in Tabs. \ref{tab:1}-\ref{tab:3}. 
The color-neutral clusters are selected from the Particle Data Group \cite{ParticleDataGroup:2022pth} summary tables for mesons in Tab. \ref{tab:2} and for baryons in Tab. \ref{tab:3}, according to the choice made in Ref. \cite{Albright:2014gva}.
We added the deuteron as a hexaquark state as well as the hypothetical sexaquark $S(uuddss)$, for which a mass of $1885$ MeV was motivated by Buccella \cite{Buccella:2020mxi} assuming a compact three-diquark structure, see also \cite{Shahrbaf:2022upc} and references therein.
The contribution of heavier clusters is discussed below. It should be mentioned that also further clusters such as penta-quarks are discussed recently
\cite{Stancu:2005jv,Maiani:2023nwj}, but those containing only three light quark flavors are not yet confirmed by experiments. 

\begin{table}
\centering
\begin{tabular}{|c|c|c|c|c|c|}
\hline
$B_i$ & colored cluster & $d_i$  & $M_i$ & $M_{{\rm th},i}^<$  & $M_{{\rm th},i}^>$  \\
 &  &  & [MeV] & [MeV] & [MeV] \\

\hline
2/3 & $D_1$ ($uu$/$ud$/$dd$) & $3$ & $783$ & $1254$ & $11.2$ \\
2/3 & $D_2$ ($us$/$ds$)& $6$  & $926$ & $1397$ & $129.6$ \\
4/3 & $4q_1$ ($D_1D_1$)& $3$ & $1095$ & $2508$ & $22.4$ \\
4/3 & $4q_2$ ($D_1D_2$)& $6$  & $1238$ & $2651$ & $140.8$ \\
5/3 & $5q_1$ ($ND_1$)& $12$  & $1251$ & $3135$ & $28.0$ \\
5/3 & $5q_2$ ($ND_2$)& $24$  & $1394$ & $3278$ & $146.4$ \\
5/3 & $5q_3$ ($\Lambda D_2$)& $48$ & $1537$ & $3421$ & $264.8$ \\
\hline
\end{tabular}
\caption{Masses of  colored multiquark clusters  according to the simple hadron mass formula (\ref{eq:masses}) together with the corresponding multiquark thresholds (\ref{eq:threshold_masses}).
$B_i$ is the baryon number and $d_i$ the degeneracy of the state $i$.
}
\label{tab:1}
\end{table}

\begin{table}
\centering
\begin{tabular}{|c|c|c|c|c|c|}
\hline
 PDG & $d_i$& $M_{\rm PDG}$ & $M_{i}$ & $M_{{\rm th},i}^<$ & $M_{{\rm th},i}^>$ \\
 mesons & & [MeV] &  [MeV] &  [MeV] & [MeV]\\
\hline
$\pi^+$/$\pi^0$ & $3$& $140$ & 140 & $1254$ & $11.2$ \\
$K^+$/$K^0$& $4$ & $494$ &  $494$ & $1397$ & $129.6$ \\
$\eta$ & $1$& $548$ & $878$ & $1349$ & $90.1$ \\
$\rho^+$/$\rho^0$& $9$  & $775$ & 783 & $1254$ & $11.2$ \\
$\omega$ & $9$& $783$ & 783 & $1254$ & $11.2$ \\
$K^{*+}$/$K^{*0}$ & $12$& $895$ & $806^{*)}$ & $2651$ & $140.8$ \\
$\eta'$ & $1$& $960$ & $878$ & $1349$ & $90.1$  \\
$a_0$ & $3$& $980$ & $1095^{*)}$ & $2508$ & $22.4$ \\
$f_0$ & $1$ & $980$ & $1095^{*)}$ & $2508$ & $22.4$ \\
$\phi$ & $3$& $1020$ & 1069 & $1540$ & $248$ \\
$h_1$ & $3$ & $1170$ & 1069 & $1540$ & $248$ \\
$a_1$/$b_1$& $18$ & $1230$ & $1095^{*)}$ & $2508$ & $22.4$ \\
$K_1(1270)$& $12$ & $1272$ &$1238^{*)}$ & $2651$ & $140.8$ \\
$f_2$ & $5$ & $1275$ & $1095^{*)}$ & $2508$ & $22.4$ \\
$f_1$ & $3$& $1282$ & $1095^{*)}$ & $2508$ & $22.4$ \\
$\eta(1295)$& $1$ & $1294$ & $1190^{*)}$ &$2603$ & $101.4$ \\
$\pi(1300)$  & $3$ & $1300$ &$1095^{*)}$ & $2508$ & $22.4$ \\
$a_2$ & $15$& $1318$ &$1095^{*)}$ & $2508$ & $22.4$  \\
$f_0(1370)$& $1$ & $1350$ &$1095^{*)}$ & $2508$ & $22.4$ \\
$\pi_1(1400)$ & $9$& $1354$ &$1095^{*)}$ & $2508$ & $22.4$ \\
$K_1(1400)$ & $12$& $1403$ &$1238^{*)}$ & $2651$ & $140.8$\\
$\eta(1405)$& $1$  & $1409$ &$1190^{*)}$ &$2603$ & $101.4$ \\
$K^*(1410)$  & $12$& $1414$ &$1238^{*)}$ & $2651$ & $140.8$\\
$\omega(1420)$ & $3$& $1425$ &$1095^{*)}$ & $2508$ & $22.4$  \\
$K_0^*(1430)$ & $4$& $1425$ &$1238^{*)}$ & $2651$ & $140.8$\\
$K_2^{*\pm}(1430)$ & $10$& $1426$ &$1238^{*)}$ & $2651$ & $140.8$ \\
$f_1(1420)$& $3$ & $1426$ &$1095^{*)}$ & $2508$ & $22.4$ \\
$K_2^{*0}(1430)$& $10$ & $1432$ &$1238^{*)}$ & $2651$ & $140.8$ \\
$\rho(1450)$  & $9$ & $1465$ & $1095^{*)}$ & $2508$ & $22.4$  \\
$a_0(1450)$ & $3$& $1474$ & $1095^{*)}$ & $2508$ & $22.4$  \\
$\eta(1475)$& $1$  & $1476$ &$1190^{*)}$ &$2603$ & $101.4$ \\
$f_0(1500)$ & $1$ & $1505$ &$1095^{*)}$ & $2508$ & $22.4$ \\
$f_2^\prime(1525)$& $5$ & $1525$ &$1095^{*)}$ & $2508$ & $22.4$ \\
$\eta_2(1645)$& $5$  & $1617$ &$1190^{*)}$ &$2603$ & $101.4$ \\
$\pi_1(1600)$  & $9$& $1662$ &$1095^{*)}$ & $2508$ & $22.4$ \\
$\omega_3(1670)$ & $7$& $1667$ &$1095^{*)}$ & $2508$ & $22.4$ \\
$\omega(1650)$ & $3$& $1670$ & $1095^{*)}$ & $2508$ & $22.4$ \\
$\pi_2(1670)$  & $15$ & $1672$ &$1095^{*)}$ & $2508$ & $22.4$ \\
$\phi(1680)$ & $3$ & $1680$ &$1381^{*)}$ & $2794$& $259.2$\\
$\rho_3(1690)$ & $21$& $1689$ &$1095^{*)}$ & $2508$ & $22.4$ \\
$K^*(1680)$  & $12$& $1717$ &$1238^{*)}$ & $2651$ & $140.8$ \\
$\rho(1700)$ & $9$& $1720$ &$1095^{*)}$ & $2508$ & $22.4$ \\
$f_0(1710)$ & $1$ & $1720$ &$1095^{*)}$ & $2508$ & $22.4$ \\
$K_2(1770)$ & $20$& $1773$ &$1238^{*)}$ & $2651$ & $140.8$ \\
$K_3^*(1780)$ & $28$& $1776$ &$1238^{*)}$ & $2651$ & $140.8$ \\
$\pi(1800)$ & $3$& $1812$ & $1095^{*)}$ & $2508$ & $22.4$ \\
$K_2(1820)$ & $20$& $1816$ &$1238^{*)}$ & $2651$ & $140.8$ \\
$\phi_3(1850)$& $7$ & $1854$ &$1381^{*)}$ & $2794$& $259.2$ \\
$\pi_2(1880)$& $15$  & $1895$ &$1095^{*)}$ & $2508$ & $22.4$ \\
$f_2(1950)$ & $5$ & $1944$ &$1095^{*)}$ & $2508$ & $22.4$ \\
$a_4(2040)$ & $27$ & $1996$ &$1095^{*)}$ & $2508$ & $22.4$ \\
$f_2(2010)$ & $5$ & $2011$ &$1095^{*)}$ & $2508$ & $22.4$ \\
$f_4(2050)$ & $9$ & $2018$ &$1095^{*)}$ & $2508$ & $22.4$ \\
$K_4^*(2045)$ & $36$& $2045$ &$1238^{*)}$ & $2651$ & $140.8$ \\
$\phi(2170)$ & $3$& $2175$ & $1381^{*)}$ & $2794$& $259.2$\\
$f_2(2300)$ & $5$& $2297$ & $1095^{*)}$ & $2508$ & $22.4$ \\
$f_2(2340)$ & $5$ & $2339$ &$1095^{*)}$ & $2508$ & $22.4$ \\
\hline
\end{tabular}
\caption{Meson masses 
and degeneracy factors $d_i$ according to the PDG \cite{ParticleDataGroup:2022pth}, 
compared to $M_i$ from 
(\ref{eq:GMOR}) for pseudoscalars or (\ref{eq:masses})
and the continuum thresholds 
(\ref{eq:threshold_masses}).
States with $^{*)}$ are understood as tetraquark states.
}
\label{tab:2}
\end{table}

\begin{table}
\centering
\begin{tabular}{|c|c|c|c|c|c|}
\hline
 PDG & $d_i$& $M_{\rm PDG}$ & $M_{i}$& $M_{{\rm th},i}^<$ & $M_{{\rm th},i}^>$  \\
 baryons &  & [MeV] & [MeV] & [MeV]& [MeV]  \\
\hline
n/p & $4$ & $939$ & $939$ & $1881$ & $16.8$ \\
$\Lambda$& $2$ & $1116$ & $1082$ & $2024$ & $135.2$ \\
$\Sigma$ & $6$& $1193$ & $1082$ & $2024$ & $135.2$ \\
$\Delta$& $16$ & $1232$ & $1251^{**)}$ & $3135$ & $28$ \\
$\Xi^0$ & $2$ & $1315$ & $1225$ & $2167$ & $253.6$ \\
$\Xi^-$ & $2$ & $1322$ & $1225$ & $2167$ & $253.6$ \\
$\Sigma(1385)$ & $6$ & $1385$ & $1394^{**)}$ & $3278$ & $146.4$ \\
$\Lambda(1405)$ & $2$ & $1405$ & $1394^{**)}$ & $3278$ & $146.4$ \\
$N(1440)$ & $4$ & $1440$ & $1251^{**)}$ & $3135$ & $28$ \\
$\Lambda(1520)$ & $4$ & $1520$ & $1394^{**)}$ & $3278$ & $146.4$ \\
$N(1520)$ & $8$ & $1520$ & $1251^{**)}$ & $3135$ & $28$ \\
$\Xi^0(1530)$ & $4$ & $1532$ & $1537^{**)}$ & $3421$ & $264.8$ \\
$\Xi^-(1530)$  & $4$ & $1535$ & $1537^{**)}$ & $3421$ & $264.8$ \\
$N(1535)$ & $4$ & $1535$ & $1251^{**)}$ & $3135$ & $28$ \\
$\Delta(1620)$ & $8$ & $1630$ & $1251^{**)}$ & $3135$ &$28$ \\
$N(1650)$ & $4$ & $1655$ & $1251^{**)}$ & $3135$ & $28$ \\
$\Sigma(1660)$ & $6$ & $1660$ & $1394^{**)}$ & $3278$ & $146.4$ \\
$\Lambda(1670)$ & $2$ & $1670$ & $1394^{**)}$ & $3278$ & $146.4$\\
$\Sigma(1670)$ & $12$ & $1670$ & $1394^{**)}$ & $3278$ & $146.4$ \\
$\Omega^-$ & $4$ & $1673$ & $1368$ & $2310$ & $372$ \\
$N(1675)$ & $12$ & $1675$ & $1251^{**)}$ & $3135$ & $28$\\
$N(1680)$ & $12$ & $1685$ & $1251^{**)}$ & $3135$ & $28$\\
$\Lambda(1690)$ & $4$ & $1690$ & $1394^{**)}$ & $3278$ & $146.4$\\
$\Xi(1690)$ & $4$ & $1690$ & $1537^{**)}$ & $3421$ & $264.8$\\
$N(1700)$ & $8$ & $1700$ & $1251^{**)}$ & $3135$ & $28$ \\
$\Delta(1700)$ & $16$ & $1700$ &$1251^{**)}$ & $3135$ & $28$\\
$N(1710)$ & $4$ & $1710$ & $1251^{**)}$ & $3135$ & $28$\\
$N(1720)$ & $8$ & $1720$ & $1251^{**)}$ & $3135$ & $28$\\
$\Sigma(1750)$ & $6$ & $1750$ &$1394^{**)}$ & $3278$ & $146.4$\\
$\Sigma(1775)$ & $18$ & $1775$ &$1394^{**)}$ & $3278$ & $146.4$\\
$\Lambda(1805)$ & $4$ & $1805$ & $1394^{**)}$ & $3278$ & $146.4$ \\
$\Xi(1820)$ & $8$ & $1823$ &$1537^{**)}$ & $3421$ & $264.8$\\
$\Lambda(1825)$ & $12$ & $1825$ &$1394^{**)}$ & $3278$ & $146.4$ \\
$N(1875)$ & $8$ & $1875$ &$1251^{**)}$ & $3135$ & $28$\\
$d$ & $3$ & $1875$ &$1251^{***)}$ & $3762$ & $33.6$\\
$S(uuddss)$ & $1$ & $1885$ &$1693^{****)}$ & $4048$ & $270.4$\\
$\Delta(1905)$ & $32$ & $1885$ & $1251^{**)}$ & $3135$ & $28$ \\
$\Lambda(1890)$ & $4$ & $1890$ & $1394^{**)}$ & $3278$ & $146.4$\\
$N(1900)$ & $8$ & $1900$ & $1251^{**)}$ & $3135$ & $28$\\
$\Sigma(1915)$ & $18$ & $1915$ &$1394^{**)}$ & $3278$ & $146.4$\\
$\Delta(1925)$ & $48$ & $1925$ & $1251^{**)}$ & $3135$ & $28$ \\
$\Sigma(1940)$& $12$ & $1940$ & $1394^{**)}$ & $3278$ & $146.4$ \\
$\Delta(1930)$ & $24$ & $1950$ & $1251^{**)}$ & $3135$ & $28$\\
$\Xi(1950)$ & $4$ & $1950$ & $1537^{**)}$ & $3421$ & $264.8$\\
$\Xi(2030)$ & $12$ & $2025$ &$1537^{**)}$ & $3421$ & $264.8$\\
$\Sigma(2030)$ & $24$ & $2030$ & $1394^{**)}$ & $3278$ & $146.4$ \\
$\Lambda(2105)$ & $14$ & $2105$ &$1394^{**)}$ & $3278$ & $146.4$ \\
$N(2195)$ & $36$ & $2220$ & $1251^{**)}$ & $3135$ & $28$\\
$\Sigma(2250)$ & $6$ & $2250$ & $1394^{**)}$ & $3278$ & $146.4$\\
$\Omega^-$(2250) & $2$ & $2252$ & $1680^{**)}$ & $3564$ & $383.2$ \\
$N(2250)$ & $20$ & $2275$ & $1251^{**)}$ & $3135$ & $28$\\
$\Lambda(2350)$ & $10$ & $2350$ &$1394^{**)}$ & $3278$ & $146.4$\\
$\Delta(2420)$ & $48$ & $2420$ &$1251^{**)}$ & $3135$ & $28$ \\
$N(2600)$ & $24$ & $2600$ & $1251^{**)}$ & $3135$ & $28$\\
\hline
\end{tabular}
\caption{Color-singlet hadrons with baryon number $B_i=1$, masses $M_{\rm PDG}$ and degeneracy factor $d_i$ according to the PDG \cite{ParticleDataGroup:2022pth}. Also shown are the masses $M_i$ according to our simple hadron mass formula (\ref{eq:masses}) and the masses of the continuum thresholds  $M_{{\rm th},i}$ for the decay into their quark constituents.
States denoted by $^{**)}$ are considered as pentaquark states; 
$^{***)}$ the deuteron is a six--quark state with molecular structure of two almost unbound nucleons;
$^{****)}$ the sexaquark is a hypothetical deeply bound
six--quark state \cite{Shahrbaf:2022upc}.
}
\label{tab:3}
\end{table}

The MHRG part takes the form of a cluster decomposition of the thermodynamic potential for quark matter 
\begin{eqnarray}
\label{eq:Omega}
\Omega_{\rm MHRG}(T,\mu,\phi,\bar{\phi}) &=& \sum_{n=2}^{N}  \Omega_n(T,\mu) + \Phi\left[\left\{S_n \right\} \right],
\\
\Omega_n(T,\mu) &=& c_n\left[{\rm Tr} \ln S_n^{-1} + {\rm Tr} (\Pi_n S_n) \right],
\label{eq:Omega-a}
\end{eqnarray}    
where $n$ denotes the total number of valence quarks and antiquarks in the cluster; $c_n=1/2$ for bosonic and $c_n=-1$ for fermionic clusters \cite{Vanderheyden:1998ph,Blaizot:2000fc}. 
The functional $\Phi\left[\left\{S_n \right\} \right]$ contains all two-cluster irreducible (2CI) closed-loop diagrams that can be formed with the complete set of cluster Green's functions $S_n$. 
We will restrict ourselves to a maximum number of $N=6$ quarks in the cluster and to the class of two-loop diagrams of the "sunset" type which are shown in Fig.~\ref{fig:sunset1} and Fig.~\ref{fig:sunset2} for the case of $N=6$.

The full quark cluster propagators $S_n$ fulfill a Dyson--Schwinger equation 
\begin{equation}
\label{eq:Dyson}
S_n^{-1} = {S_n^{(0)}}^{-1} - \Pi_n~,~n=2, \dots, N,
\end{equation}    
where $S_n^{(0)}$ is the free $n-$quark cluster propagator and the cluster selfenergy $\Pi_n$ should be obtained from
\begin{equation}
\label{eq:Pi}
\Pi_n= \frac{\partial \Phi}{\partial S_n}~.
\end{equation}

The  free $n-$quark cluster propagator for $n=2, \dots, N$ in general is obtained from the free quark propagator by a product ansatz and subsequent summation over $n-1$ Matsubara frequencies in order to obtain a one-frequency function, see, e.g., Ref.~\cite{Bastian:2018wfl}. 
Here we perform a cluster virial expansion which we restrict to the second virial coefficients for interactions of pairs of clusters, including also quark pairs as the simplest example.  
In that case we obtain the  free $a-$quark cluster propagator by performing one Matsubara summation in the product ansatz for the bipartition of the cluster as illustrated in Appendix \ref{app:1}. 
The generalized Matsubara frequencies which include the coupling to the Polyakov-gauge gluon background field, are given in Tab.~\ref{tab:Matsubara}.

\begin{figure}[!htb]
\includegraphics[width=0.7\linewidth]{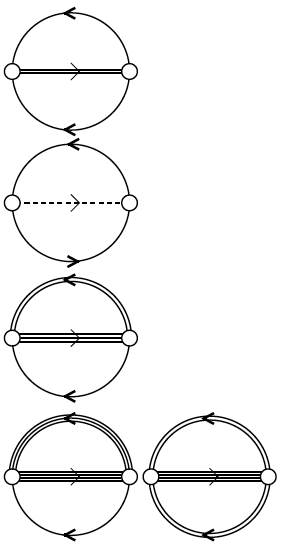}
\caption{
Feynman diagrams for the $\Phi$ functional in the cluster two-loop approximation. 
A $n$-quark cluster Green's function is shown as $n$ solid lines bound with one arrow, a meson Green's function is a dashed line with an arrow. 
Starting from the top, the figure shows the diquark as two-quark state, the meson as quark-antiquark state, the nucleon as quark-diquark state and the 4--quark cluster as nucleon-quark or diquark-diquark state.
}
\label{fig:sunset1}
\end{figure}

\begin{figure}[!htb]
\includegraphics[width=\linewidth]{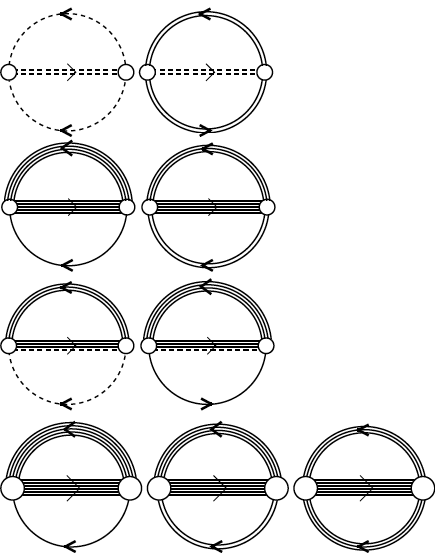}
\caption{
Continuation of Fig. \ref{fig:sunset1}. Starting from the top, we show the Feynman diagrams for the $\Phi$ functional in the cluster two-loop approximation whith the tetraquark as meson-meson or diquark-antidiquark state, the 5--quark cluster as 4--quark-quark state or nucleon-diquark state, the pentaquark as nucleon-meson state or 4--quark-antiquark state and the hexaquark channels: 5--quark-quark, 4--quark-diquark and nucleon-nucleon states.
}
\label{fig:sunset2}
\end{figure}

\begin{table*}
\centering
\begin{tabular}{|c|c|c|c|c|c|c|c|}
\hline
symbol & cluster & quarks & $n$ & $a$ & bipartition & Matsubara frequency & color \\
$i$ & & (antiquarks) & & & & $\omega_n^{(a)}$ & \\
\hline
q      &  quark  & 1 (0) & 1 & 1 & --    & $((2n+1)\pi T -i\mu)\mathbbm{1}_c -  A_4 $    & triplet \\
D      & diquark & 2 (0) & 2 & 2& (q+q)   & $(2n\pi T - i 2\mu)\mathbbm{1}_c +  A_4 $     & antitriplet \\
M      &  meson  & 1 (1) & 2 & 0& (q+$\mathrm{\bar{q}}$)  & $(2n\pi T)\mathbbm{1}_c $ & singlet \\
N      & nucleon & 3 (0) & 3 & 3 & (q+D)   & $((2n+1)\pi T - i 3\mu)\mathbbm{1}_c  $   & singlet \\
F      & 4-quark & 4 (0) & 4 & 4 & (q+N), (D+D)    & $(2n\pi T - i 4\mu)\mathbbm{1}_c -  A_4 $ & triplet \\
T      &tetraquark& 2 (2) & 4 & 0 & (M+M), (D+$\mathrm{\bar{D}}$)    & $(2n\pi T)\mathbbm{1}_c$  & singlet \\
Q      & 5-quark & 5 (0) & 5 & 5 & (q+F), (D+N)   & $((2n+1)\pi T - i 5\mu)\mathbbm{1}_c +  A_4 $ & antitriplet \\
P      &pentaquark& 4 (1) & 5 & 3 & (M+N), ($\mathrm{\bar{q}}$+F)   & $((2n+1)\pi T - i 3\mu)\mathbbm{1}_c $ & singlet \\
H      &hexaquark& 6 (0) & 6 & 6 & (q+Q), (D+F), (N+N)   & $(2n\pi T - i 6\mu)\mathbbm{1}_c $    & singlet \\
\hline
\end{tabular}
\caption{Matsubara frequencies for an $n-$particle cluster with $a$ net valence quarks in the Polyakov-gauge quark model, obtained as composite of a pair of lower order clusters, see Fig.~\ref{fig:sunset1} and Fig.~\ref{fig:sunset2}}
\label{tab:Matsubara}
\end{table*}

The equations (\ref{eq:Omega}),  (\ref{eq:Omega-a}),  (\ref{eq:Dyson}) and  (\ref{eq:Pi}) form a closed set of equations that for its solution requires the knowledge of the free $n-$quark cluster propagator $S_n^{(0)}$ and the choice of the 2CI set of diagrams of the $\Phi-$functional, built with the set of full cluster propagators $\left\{S_n\right\}$.
Following the spirit of Refs.~\cite{Vanderheyden:1998ph,Blaizot:2000fc}, 
we consider here the quark density and entropy density as first derivatives of the thermodynamic potential with respect to the quark chemical potential $\mu$ and the temperature $T$, respectively.
From the cluster decomposition (\ref{eq:Omega}) of the  MHRG thermodynamic potential follows the cluster decomposition of the quark density

\begin{eqnarray}
\label{eq:n}
n_{\rm MHRG}(T,\mu)&=& 
-\frac{\partial \Omega_{\rm MHRG}(T,\mu)}{\partial \mu} 
\nonumber\\
&=&\sum_i a_i \, n_i(T,\mu) ,
\nonumber\\
\end{eqnarray} 
where $a_i$ is the net quark number in the $n_i-$particle state with the partial density defined as
\begin{eqnarray}
n_i(T,\mu) 
&=& -\frac{\partial}{\partial \mu}
\left\{
d_i \int\frac{d^3 q}{(2\pi)^3}\int\frac{d\omega}{2\pi}
\left[ \ln \left(-S_{n_i}^{-1}\right)+\right.\right.
\nonumber\\
&&\left.\left.+\operatorname{Tr} \left(\Pi_{n_i}~S_{n_i} \right) \right]
+\sum_{\stackrel{i_1,i_2}{i_1+i_2={n_i}}}\Phi[S_{i_1},S_{i_2},S_{n_i}]
\right\}
\nonumber\\
\label{eq:n_a}
\end{eqnarray} 
and $d_i$ denotes the degeneracy factor for $n_i-$particle states.
Using the spectral decomposition rule for an analytic complex function $F(\omega)$ 
\begin{eqnarray}
F(i\omega_n) = \int_{-\infty}^\infty\frac{d\omega}{2\pi}\frac{2\, \mathrm{Im} F(\omega)}{\omega - i\omega_n}
\end{eqnarray}
and the Matsubara summation in the case of many-quark Green's functions coupled to the gluon background field
(see Tab. \ref{tab:Matsubara} and Appendix \ref{app:1-3})
\begin{eqnarray}
\frac{1}{3}\sum_{c=1}^3 \frac{1}{\beta}\sum_{n} 
\frac{1}{i\omega_n^{(a)}-\omega \mathbbm{1}_c} = f^{(a),+}_\phi(\omega)
\end{eqnarray}
with the relation $\partial f^{(a)}(\omega)/\partial \mu = -\partial f^{(a)}(\omega)/\partial \omega$, we get for Eq. (\ref{eq:n_a}) now
\begin{eqnarray}
n_i(T,\mu) &=& - d_i c_{a_i}\int\frac{d^3 q}{(2\pi)^3}\int\frac{d\omega}{\pi}f^{(a_i),+}_\phi(\omega)
\frac{\partial}{\partial \omega}\left[{\rm Im} \ln \left(-S_{n_i}^{-1}\right) \right.
\nonumber\\
&&+\left. {\rm Im} \left(\Pi_{n_i}~S_{n_i} \right) \right]
+\sum_{\stackrel{i_1,i_2}{i_1+i_2=n_i}}\frac{\partial \Phi[S_{i_1},S_{i_2},S_{n_i}]}{\partial \mu}~, 
\label{eq:n_a_} 
\end{eqnarray} 
where a partial integration over $\omega$ has been performed. 
For two-loop diagrams of the sunset type holds a cancellation
\cite{Vanderheyden:1998ph,Blaizot:2000fc}, which we generalize here for cluster states
\begin{eqnarray}
d_i c_{a_i}\int\frac{d^3 q}{(2\pi)^3}\int\frac{d\omega}{\pi}&&
f^{(a_i),+}_\phi(\omega)
\frac{\partial}{\partial \omega}\left({\rm Re}\Pi_{n_i}~ {\rm Im}S_{n_i} \right)\nonumber\\
&-&\sum_{\stackrel{i_1,i_2}{i+j=n_i}}\frac{\partial \Phi[S_{i_1},S_{i_2},S_{n_i}]}{\partial \mu} =0 ~.
\nonumber\\
\label{cancel}
\end{eqnarray} 
Using generalized optical theorems 
\cite{Bastian:2018mmc}
and the polar representation of the $n$--quark cluster Green's function $S_n=|S_n| \exp(i\delta_n)$, where $\delta_n$ is the cluster phase shift, one obtains
\begin{eqnarray}
\frac{\partial}{\partial \omega}&&\left[{\rm Im} \ln \left(-S_n^{-1}\right) 
+ {\rm Im}\Pi_n~ {\rm Re} S_n \right] 
\nonumber\\
&=&2\, {\rm Im} \left[S_n~{\rm Im}\Pi_n~\frac{\partial}{\partial \omega} S_n^*~{\rm Im}\Pi_n  \right]\nonumber\\
&=&-2\sin^2\delta_n \frac{\partial \delta_n}{\partial \omega}.
\end{eqnarray}
The density is obtained in the form of a generalized 
Beth-Uhlenbeck EoS
\begin{eqnarray}
\label{eq:n}
n_{\rm MHRG}(T,\mu)&=& \sum_i a_i \, d_i\, c_{a_i}\int \frac{d^3q}{(2\pi)^3} 
\int_0^\infty \frac{d\omega}{\pi}
\left\{f^{(a_i),+}_\phi \right.
\nonumber\\
&&\left.-\left[f^{(a_i),-}_\phi\right]^*\right\}
2 \sin ^2 \delta_{n_i}(\omega,q) \frac{\partial  \delta_{n_i}(\omega,q)}{\partial \omega} ~,
\nonumber\\
\end{eqnarray}
where the properties of the distribution function $f^{(a),+}_\phi$ and the phase shift with respect to 
reflection $\omega \to -\omega$ have been used and the 
"no sea" approximation has been employed which removes the divergent vacuum contribution.
The Polyakov--loop modified distribution functions are defined as
\begin{eqnarray}
\label{eq:PL-function}
    f^{(a),\pm}_{\phi}~\stackrel{\text{(a even)}}{=}&&\frac{({\phi} - 2\bar{\phi} y_a^\pm) y^\pm_a + {y_a^\pm}^3}{1 - 3 ({\phi} - \bar{\phi} y_a^\pm) y_a^\pm - {y_a^\pm}^3}~,\\
    f^{(a),\pm}_{\phi}~\stackrel{\text{(a odd)}}{=}&&\frac{(\bar{\phi} + 2\phi y_a^\pm) y^\pm_a + {y_a^\pm}^3}{1 + 3 (\bar{\phi} + \phi y_a^\pm) y_a^\pm + {y_a^\pm}^3}~,
\end{eqnarray}
where $y^\pm_a=e^{-\left( \omega \mp a\mu\right)/T}$ 
and $a$ is the net number of valence quarks present in the cluster. 
See Appendix \ref{app:1} for a detailed derivation.

In an analogous manner follows for the MHRG entropy density
\begin{eqnarray}
s_{\rm MHRG}(T,\mu)&=& -\frac{\partial \Omega}{\partial T} 
= \sum_i s_i(T,\mu) \nonumber\\
&=& \sum_i  d_i \, c_{a_i}\int \frac{d^3q}{(2\pi)^3}\int \frac{d\omega}{\pi}
\left\{
\sigma^{(a_i),+}_\phi \right.
\nonumber\\
&&\left.
+\left[\sigma^{(a_i),-}_\phi\right]^*
\right\}
 2 \sin ^2 \delta_{n_i}(\omega,q) \frac{\partial  \delta_{n_i}(\omega,q)}{\partial \omega} 
~,
\nonumber\\
\label{eq:s}
\end{eqnarray} 
where $\sigma^{(a)} =  f^{(a)}_\phi  \ln f^{(a)}_\phi (-)^a [1(-)^a f^{(a)}_\phi] \ln [1(-)^a f^{(a)}_\phi]$ and $f^{(a)}_\phi$ is the cluster distribution function for a net quark number $a$ modified by the traced Polyakov loop.

The Eqs. (\ref{eq:n}) and (\ref{eq:s}) are consistent with each other because they follow from the functional for the thermodynamic potential (\ref{eq:Omega}) in the case of two-loop skeleton diagrams for the $\Phi$- functional \cite{Vanderheyden:1998ph,Blaizot:2000fc}, where it has been shown that the correlation contribution vanishes, see Eq. (\ref{cancel}).
The formula for the pressure as thermodynamical potential can be obtained from Eq. (\ref{eq:n}) by integration  over the quark chemical potential $\mu$.
Analogously, it can be obtained from Eq. (\ref{eq:s}) by integration over $T$
\begin{equation}
    \label{eq:pressure}
    p(T,\mu) = \int_0^T dT' s(T',\mu)~.
\end{equation}

It is instructive to consider the two limits $\phi=\bar{\phi}=1$ (deconfinement and color singlet states) and $\phi=\bar{\phi}=0$ (confinement).
In the situation where $\phi=\bar{\phi}=1$, the distribution functions reduce to
\begin{eqnarray}
f^{(a=0,2,4,\dots),\pm}_{\phi=1}&=&
\frac{y_a^\pm}{1-y_a^\pm}~,\\
f^{(a=1,3,5,\dots),\pm}_{\phi=1}&=&
\frac{y_a^\pm}{1+y_a^\pm}~,
\end{eqnarray}
corresponding to the ordinary Bose (Fermi) distribution functions for even (odd) numbers $a$ of quarks in the cluster. 
In the case of confinement when $\phi=\bar{\phi}=0$, the 
distribution functions take the form
\begin{eqnarray}
\label{eq:f-conf}
f^{(a=0,2,4,\dots),\pm}_{\phi=0}&=&
\frac{{y_a^\pm}^3}{1-{y_a^\pm}^3}~,\\
f^{(a=1,3,5,\dots),\pm}_{\phi=0}&=&
\frac{{y_a^\pm}^3}{1+{y_a^\pm}^3}~.
\end{eqnarray}
From Eq. (\ref{eq:f-conf}) one can see how the Polyakov-loop suppression of colored states acts. 
Since ${y_a^\pm}^3(T)=y_a^\pm(T/3)$, in the situation of confinement when $\phi=0$, the generalized distribution functions of odd (even) colored clusters become Fermi (Bose) functions with the temperature $T$ being replaced by $T/3$, so that in this case of full Polyakov-loop suppression the thermal occupation is the same as for the unsuppressed case at one third of the temperature.

We bring expressions (\ref{eq:n})  in another form where we split the extended phase shift $\delta_i(\omega, q)$ into its bound state part with a step-up at $\omega=E_i(q)$ and the part related to the elastic scattering phase shift $\delta_i^{\rm sc}(\omega, q)$ depending on the energy $\omega$ of relative motion of the subclusters and the total momentum $q$ of the cluster. 
The net baryon density (\ref{eq:n}) is the difference of contributions from particles and antiparticles,
\begin{eqnarray}
\label{eq:n+n-}
n_{\rm MHRG}(T,\mu)&=&n^+_{\rm MHRG}(T,\mu)-n^-_{\rm MHRG}(T,\mu)~.
\end{eqnarray}
Applying integration by parts to the Beth-Uhlenbeck equation and considering only the color singlet hadrons for which $f^{(a),+}_{\phi}=f^{(a),+}_{\phi=1}$, the particle contribution can be brought to the form 
\begin{eqnarray}
\label{eq:n+}
n^+_{\rm MHRG}(T,\mu)&=& \sum_i a_i \, d_i\, c_{a_i} \int \frac{d^3q}{(2\pi)^3} 
\left\{f^{(a),+}_{\phi=1}(E_i(q))
\right.\nonumber \\
&& \left.-f^{(a),+}_{\phi=1}(E_{{\rm thr},i})
+ \int \frac{d\omega}{\pi\,T}
f^{(a),+}_{\phi=1}\left[\delta^{\rm sc}_i(\omega,q)-
\right.\right.\nonumber \\ && \left.\left.
\frac{1}{2} \sin (2 \delta^{\rm sc}_i(\omega,q)) \right]\right\} ~,
\end{eqnarray}
and a similar relation for the antiparticles.
Here, the bound state part can be addressed to the first two contribution, similar to the Planck-Larkin-Brillouin partition function for the hydrogen plasma
\cite{Kraeft:1986}.
Within the generalized cluster Beth-Uhlenbeck formula,
we have for the $a$-quark correlation the continuum contribution as expressed by the phase shifts, and, in addition, the contribution of bound states, if they exist. 
For the entropy density (\ref{eq:s}), a similar decomposition as (\ref{eq:n+n-}) and (\ref{eq:n+}) can be made.

As well known, the subdivision of contribution from bound states and from the continuum of scattering states is arbitrary. Because of the Levinson theorem we can introduce an extended scattering phase shift which jumps by $\pi$ for every bound state at the bound state energy which is below the continuum edge of scattering states (so-called "step-up" phase shift).

\section{Exploratory model calculation for multiquark correlations}
\label{sec:model}

\subsection{Quark masses and continuum thresholds}
\label{ssec:quarkmass}

The temperature and chemical potential dependent quark masses determine the behavior of the continuum thresholds. 
We want to point out that in non-confining NJL models a smooth temperature dependence of quark masses has been obtained which results from the consideration of a medium of uncorrelated quarks in these models. 
Because in reality the quarks are strongly correlated (confined) inside hadrons even for temperatures up to the chiral crossover $T_c(\mu)$, any medium modification of quark masses shall be strongly suppressed. 

Improved microscopic calculations within a rank-2 separable Dyson-Schwinger equation (DSE) model \cite{Blaschke:1999ab} later extended within the Polyakov-loop DSE model \cite{Horvatic:2010md,Benic:2012ec}, and recent results within the confining density functional approach \cite{Ivanytskyi:2022oxv} motivate a steep
medium dependence (switch) from a large constituent quark mass in cool and dilute matter to the current quark mass of the QCD Lagrangian in a hot and dense medium.

\begin{figure}[!htb]
\begin{center}
\includegraphics[width=0.5\textwidth]{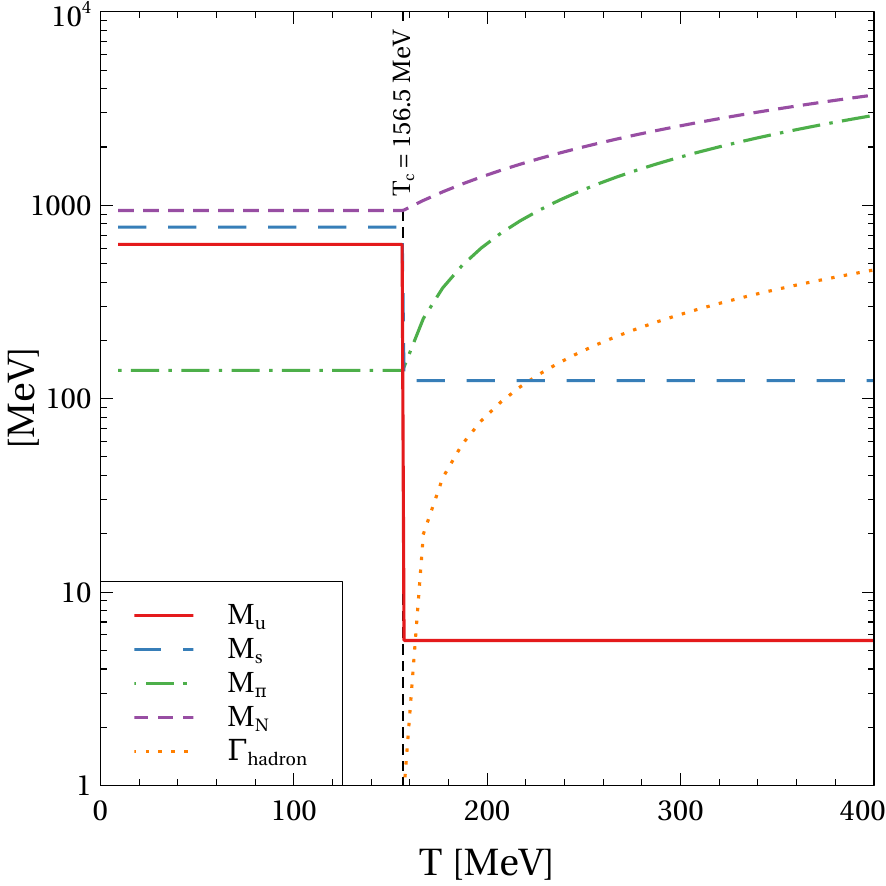}
\end{center}
\caption{Mass spectrum of light and strange quarks used in this work, together with mass and decay width of the Breit-Wigner model for the pion and the nucleon as generic examples for hadrons as a function of temperature for vanishing chemical potential $\mu/T=0$.}
\label{fig:masses}
\end{figure}

Therefore, we assume in the present work a "sudden switch" scenario for the quark masses.
It consists of a constant and large quark mass as a reflection of the quark confinement within the hadronic phase of the model that is delimited by the Mott dissociation temperature of the hadrons
(and multi-quark states) $T_{\rm Mott}(\mu)$, and a tail function for the medium-dependent quark mass in the QGP phase, which for simplicity we assume here to coincide with the constant current quark masses $m_f$,
\begin{eqnarray}
    \label{eq:quark_mass}
    M_f(T,\mu)&=&M_f\Theta(T_{\rm Mott}(\mu)-T)
    \nonumber\\
    &&+m_f\Theta(T-T_{\rm Mott}(\mu)).
\end{eqnarray} 
Here the subscript $f=q,s$ denotes the light ($q=u,d$) and strange ($f=s$) quark flavors with their corresponding current quark masses $m_f$ and the constituent quark masses $M_f=M_f(0,0)$ in the vacuum.
This quark mass model is depicted for light quarks as the red solid line in Fig. \ref{fig:masses}.

We are aware that such a "sudden switch" model is a strong idealization of the reality and could lead to a kinky behaviour of the resulting thermodynamic functions, in particular when they are derivatives of the thermodynamic potential.

Consistent with the quark confinement property we make in our model
the idealizing assumption of constant hadron masses. 
This captures, however, the fact that hadrons are almost pointlike, strongly localized states which "feel" the medium only when it provides conditions very close to the Mott dissociation.
In this case, hadron-hadron interactions induce finite mass shifts and collisional broadening.

\begin{figure}[!htb]
\begin{center}
\includegraphics[width=0.5\textwidth]{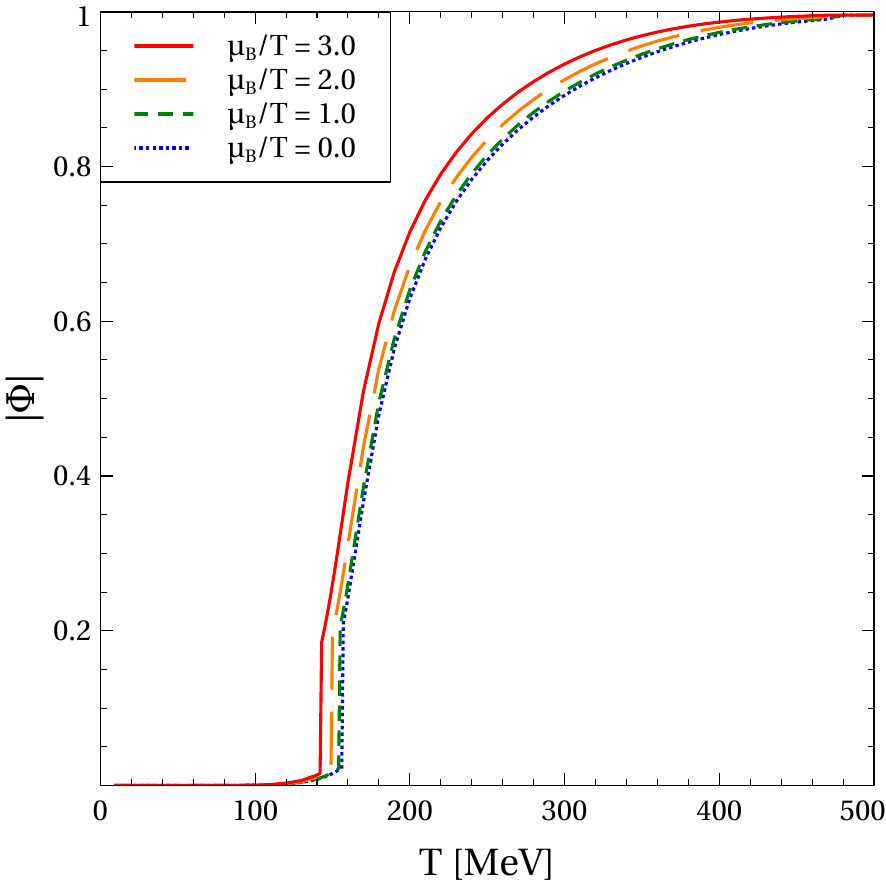}
\end{center}
\caption{Temperature dependence of the Polyakov loop absolute value calculated for several values of $\mu_B/T$ indicated in the legend.}
\label{fig:PL}
\end{figure}

We extend the formula (\ref{eq:quark_mass}) to finite $\mu_B$ by using the chemical potential dependence the pseudo-critical temperature $T_{c}(\mu)$ which resulted from the fitting the lattice QCD results obtained by Taylor expansion techniques for small chemical potentials \cite{Bazavov:2018mes} also for estimating the Mott temperature,
\begin{equation}
\label{eq:Tc-mu}
    T_{\rm Mott}(\mu)=T_{\rm Mott}(0)\left[1-\kappa\left(\frac{\mu}{T_{\rm Mott}(0)}\right)^2\right].
\end{equation}
We use the values $\kappa=0.012$ and $T_{\rm Mott}(0)=156.5$ MeV from the fit of the $\mu$- dependence of the pseudocritical temperature 
$T_c(\mu)=T_{\rm Mott}(\mu)$ found in \cite{Bazavov:2018mes}, restricting ourselves to the $\mathcal{O}(\mu_B^2)$ part of the expansion and neglecting the $\mathcal{O}(\mu_B^4)$ term because of the smallness of the associated 
$\kappa_4$ parameter found in that work. 

With the medium dependence of the quark mass being defined, one can solve the gap equation for obtaining the value of the traced Polyakov loop which would correspond to an extremum of the thermodynamic potential
(\ref{eq:omega_total}),
\begin{equation}
    \label{eq:PL-gap}
    \frac{\partial \Omega}{\partial \phi}=0~.
\end{equation}
This can be done first at the quark mean field level, including the perturbative QCD contribution. Then, the effect of colored clusters can be included, which are depending on $\phi$ and $\bar\phi$ via the Polyakov-loop generalized multiquark distribution functions 
(\ref{eq:fa-phi-even}) and (\ref{eq:fa-phi-odd}).
In Fig. \ref{fig:PL}, we show the temperature dependence of the absolute value of the Polyakov loop for different values of chemical potentials $\mu_B/T$.
These results include the backreaction from colored clusters which turn out to be negligibly small.
For temperatures below $T_{\rm Mott}(\mu)$, because of the large quark mass, the values of the traced Polyakov loop stay close to zero, reflecting confinement. 
Also in the transition region above the Mott temperature, up to $T\sim 200$ MeV, this order parameter for deconfinement stays below $|\phi|\sim 0.5$, indicating strong color correlations induced by the gluon background field. It can also be seen from Fig. \ref{fig:PL} that for temperatures above $T\sim 350$ MeV,
the order parameter asymptotically approaches 
$|\phi|= 1.0$, the case of deconfinement.

\subsection{Hadron mass formulas}
\label{ssec:hadronmass}

\subsubsection{Multiquark clusters}
In the model exploited in the present work, we do not obtain the hadron masses
as solutions of 
the corresponding equations of motion. 
Instead, we assume that below the Mott dissociation temperature 
$T_{\rm Mott}(\mu)$ the hadron masses $M_i$ are medium independent and follow the approximate relation \cite{Jankowski:2012ms}
\begin{equation}
\label{eq:masses}
M_i=(N_q + N_{\bar q})M_q + (N_s + N_{\bar s})M_s - (n-1) B ~,
\end{equation}
where $M_{q(s)}=627$ (770) MeV is the light (strange) quark mass in vacuum and $B=471$ MeV is the binding energy per quark bond, i.e. per relative Jacobi coordinate of which we have $n-1$ in a multiquark state composed of $N_q$ ($N_s$) light (strange) quarks and 
$N_{\bar q}$ ($N_{\bar s}$) light (strange) antiquarks. 
Here, $n=N_q + N_{\bar q} + N_s + N_{\bar s}$ is the total number of quarks and antiquarks and $a=N_q + N_s - N_{\bar q} - N_{\bar s}$ is the net quark number in the multiquark state (hadron).

From the medium-dependent quark mass spectrum follows immediately the temperature and chemical potential dependent threshold mass of the continuum states for a multiquark bound state species $i$
\begin{eqnarray} 
\label{eq:threshold_masses}
    M_{\rm thr,i}(T,\mu)&=&M_{\rm thr,i}^<\Theta(T_{\rm Mott}(\mu)-T)
    \nonumber\\
    &&+M_{\rm thr,i}^>\Theta(T-T_{\rm Mott}(\mu))~,\\
\label{eq:Mth<}
    M_{\rm thr,i}^<&=& (N_q + N_{\bar q})M_q + (N_s + N_{\bar s})M_s,\\
    M_{\rm thr,i}^>&=&(N_q + N_{\bar q})m_q + (N_s + N_{\bar s})m_s.
\label{eq:Mth>}
\end{eqnarray}
This behavior of the threshold masses entails that all PDG hadrons undergo the Mott transition from bound states below $T_{\rm Mott}(\mu)$ to unbound correlations in the continuum above that temperature, see Tabs. 
\ref{tab:2} and \ref{tab:3}.

For temperatures above $T_{\rm Mott}$, a linear temperature dependence of the mass and the decay width of the resonance is adopted
\begin{eqnarray}
\label{eq:mass}
M_i(T)&=&M_{i,0}+\alpha_1\, \Gamma_i(T),\\
\Gamma_i(T)&=&\alpha_2\, (T-T_{\rm Mott})\,
\Theta(T-T_{\rm Mott})~,
\label{eq:width}
\end{eqnarray}
where $M_{i,0}$ stands for the vacuum mass of the resonance $i$ that is 
taken from the particle data tables \cite{ParticleDataGroup:2022pth} and remains
unchanged for all temperatures below $T_{\rm Mott}$.
Since the Mott temperature is identified with the pseudocritical temperature $T_c$ obtained from lattice QCD simulations \cite{Bazavov:2018mes}, the ansatz (\ref{eq:mass}), (\ref{eq:width}) leaves us with only two free 
parameters, $\alpha_1=11.4$ and $\alpha_2=1.9$, which 
determine the slopes of masses and widths, respectively. 
Their values are determined such that the behaviour of the total entropy density and the total baryon density at $T_c$ are continuous.
This behavior is illustrated in Fig.~\ref{fig:masses} for the nucleon and the pion.

For the higher-lying meson (baryon) states which all decay into their ground state and a meson (mostly pion), we adopt a mass formula that assumes a tetraquark (pentaquark) structure, i.e. adding $2M_q-2B=312$ MeV to the ground state mass. The continuum threshold for these states is then 
$2M_q=1254$ MeV ($2m_q=11.2$ MeV) above that for the corresponding ground state for $T<T_{\rm Mott}$ 
($T>T_{\rm Mott}$).
While in our calculations the color singlet hadron masses are taken from the PDG \cite{ParticleDataGroup:2022pth}, 
the masses of the colored clusters and the continuum thresholds require an underlying quark model, see tables \ref{tab:1} - \ref{tab:3}. 

\subsubsection{Pseudoscalar meson masses}
The pseudoscalar mesons play a special role as the pseudo-Goldstone bosons of the broken chiral symmetry in low-energy QCD. Therefore, their masses follow a Gell-Mann--Oakes--Renner relation
\begin{equation}
    \label{eq:GMOR}
    M_P^2 = - \frac{(N_qm_q+N_sm_s)(N_q\langle \bar q q\rangle + N_s\langle \bar s s\rangle)}{(N_q-N_s)f_\pi^2 + 2N_s f_K^2},
\end{equation}
where for the $P= \pi \,(K)$, the pion (kaon) mass of 
$140$ ($494$) MeV is obtained with
$N_q=1 (1)$, $N_s=0 (1)$, $m_q=5.6$ MeV, $m_s=124$ MeV, 
$\langle \bar q q\rangle = - (242~{\rm MeV})^3$, 
$\langle \bar s s\rangle = - (245~{\rm MeV})^3$, $f_\pi=92.4$ MeV and $f_K=113$ MeV.

\subsubsection{$\eta$ mesons}
For the ground state $\eta$ mesons, we postulate a mass formula composed as a weighted sum of light and strange 
quark masses with the average mesonic binding energy 
\begin{equation}
    M_\eta=M_{\eta^\prime}=\frac{2}{3}M_s+\frac{4}{3}M_q-B,
\end{equation}
with the corresponding threshold masses 
\begin{eqnarray} 
\label{eq:threshold_eta}
    M_{\rm thr,\eta}(T,\mu)&=&\left(\frac{2}{3}M_s+\frac{4}{3}M_q\right)\Theta(T_{\rm Mott}(\mu)-T)
    \nonumber\\
    &+&\left(\frac{2}{3}m_s+\frac{4}{3}m_q\right)\Theta(T-T_{\rm Mott}(\mu)).
\end{eqnarray}
The higher lying $\eta$ states are understood as tetraquarks, i.e. adding a light quark-antiquark pair as in the case of the higher lying meson and baryon states.
\begin{figure}[!thb]
    \centering
    \includegraphics[width=0.85\columnwidth, height=0.8 \columnwidth]{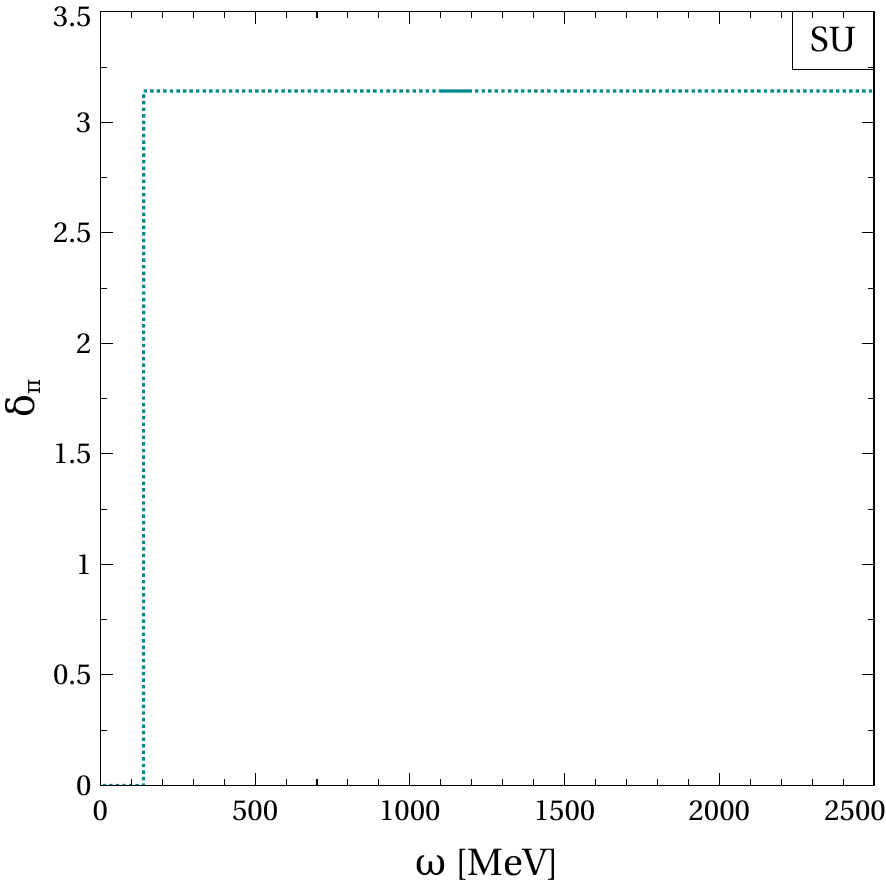}
    \includegraphics[width=0.85\columnwidth, height=0.8 \columnwidth]{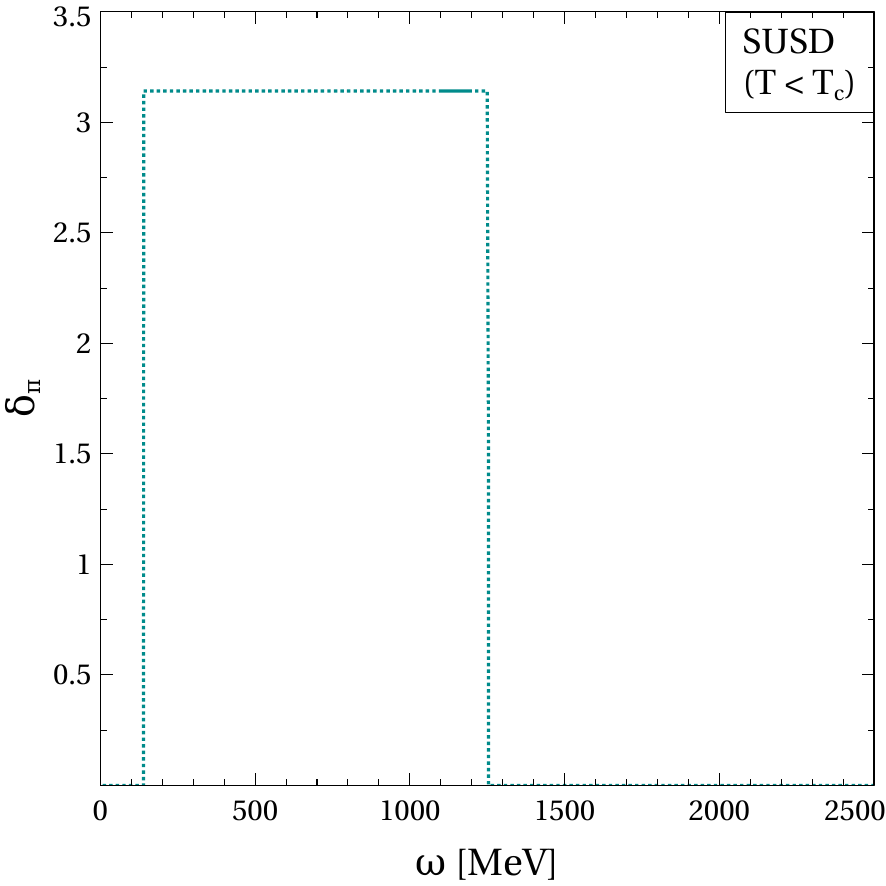}
    \includegraphics[width=0.85\columnwidth, height=0.8 \columnwidth]{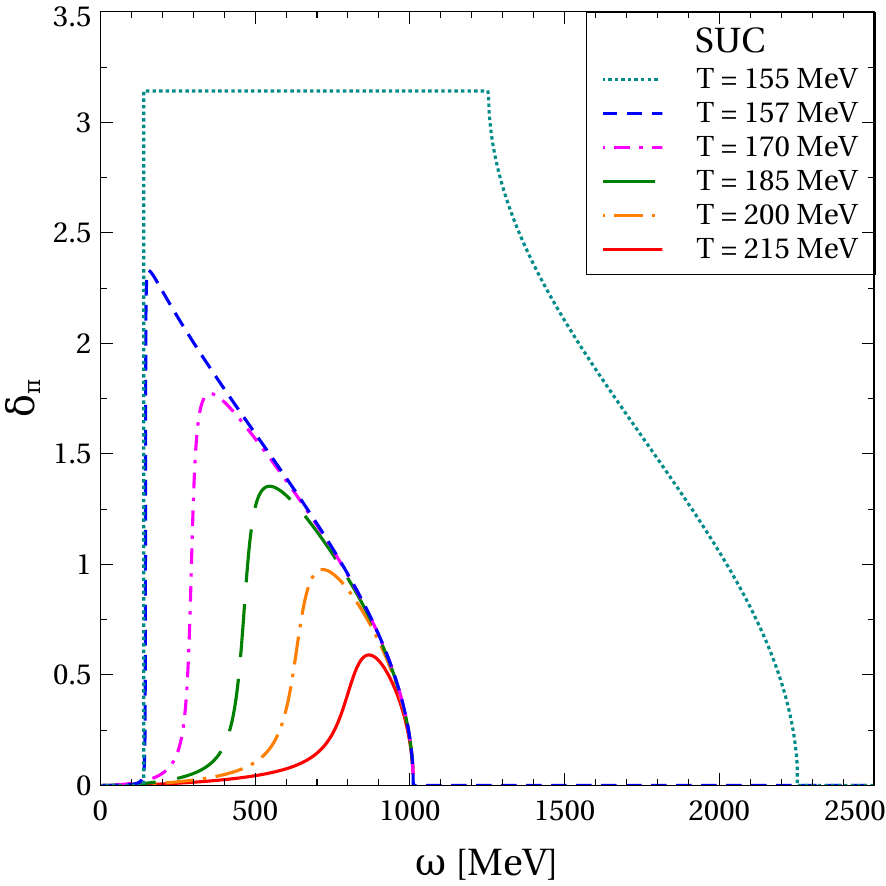}
    \caption{
    Approximations for the hadronic phase shift for the example of the pion at rest in the medium. Upper panel: step-up (SU),  middle panel: step-up-step-down (SUSD), bottom panel: step-up-continuum (SUC) models.}
    \label{fig:phase_shift_models}
\end{figure}

\subsection{Phase shift model}

In this work, we want to introduce a generic model ansatz for the phase shift $\delta_i(M,T,\mu)$ that shares all essential features with results that have been obtained, e.g., within NJL model calculations for the case of the pion \cite{Dubinin:2013yga,Blaschke:2013zaa} 
and within the PNJL model for the pion and the diquark \cite{Blaschke:2014zsa}
in quark matter.

For the extended phase shifts, which include the bound state contribution as "step-up" by $\pi$, we can use experimental determined ones, including the binding energies, or results from a potential model calculation. 

\begin{figure}[!htb]
\begin{center}
\includegraphics[width=\columnwidth]{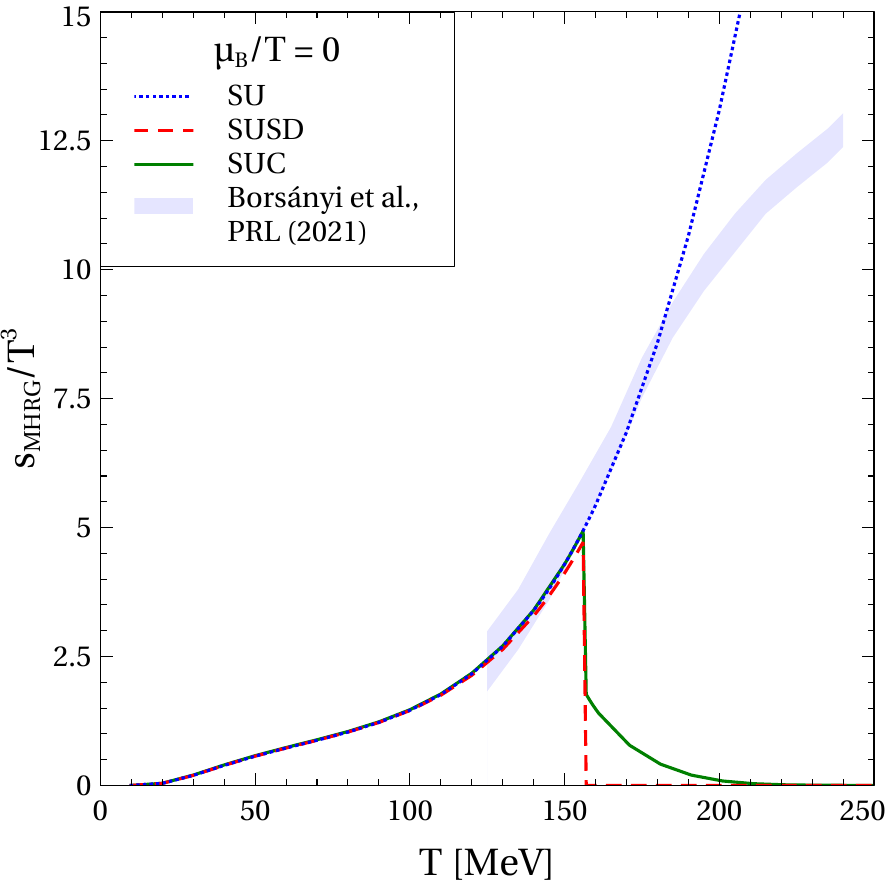}
\includegraphics[width=\columnwidth]{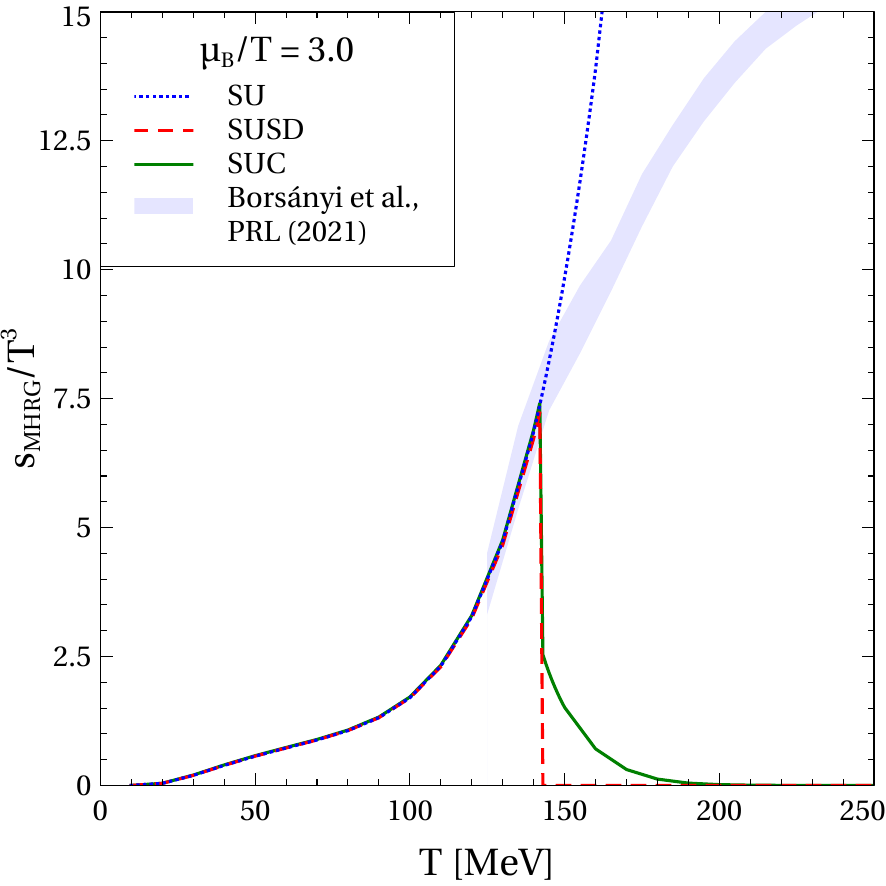}
\end{center}
\caption{Scaled entropy density of hadrons $s_{\rm MHRG}/T^3$ as a function of temperature $T$ calculated for $\mu_B/T=0$ (upper panel) and $\mu_B/T=3.0$ (lower panel) calculated for the SU (blue dotted curves), SUSD (red dashed curves) and SUC (green solid curves) models of the hadronic phase shifts discussed in the text. 
The shaded regions correspond to the lattice QCD results \cite{Borsanyi:2021sxv}, while the thick solid line represents calculations within the HRG, see \cite{Albright:2014gva}.}
\label{fig:MHRG}
\end{figure}

Three simple approximations 
which are illustrated in the three panels of 
Fig.~\ref{fig:phase_shift_models}
shall be considered: 
\begin{enumerate}
\item  The phase shifts jump up from zero to $\pi$ at the bound state mass and then remain constant.
This simplest step-up model is labelled "SU" and shown in the upper panel of
Fig. \ref{fig:phase_shift_models}.
For the full set of hadronic states in the tables \ref{tab:2}, \ref{tab:2} extracted from the listing of the PDG \cite{ParticleDataGroup:2022pth}, this 
model reproduces the HRG thermodynamics, see the dotted lines in Fig. \ref{fig:MHRG}.  
The SU model violates the Levinson theorem and does not account for the dissociation of hadrons at high temperatures and densities. 
It overestimates the contribution of clusters.
\item 
The simplest way to account not only for the formation of bound states but also for their dissociation by simultaneously fulfilling the Levinson theorem is to add a step down by $\pi$ at $E_{{\rm thr},i}$ to the step-up of the phase shift by $\pi$ at the bound state energy $E_i$. 
In this approximation, labelled "SUSD" in the middle panel of Fig. \ref{fig:phase_shift_models}, the $\omega$-integral in the Beth-Uhlenbeck 
equation (\ref{eq:n+}) for the hadron contribution to the net baryon density can be trivially performed and results in the absence of this contribution for temperatures $T>T_{\rm Mott}$. 
By the definition of this SUSD model, resonances in the continuum are absent.
In this case the entropy of the MHRG model which follows from Eq. (\ref{eq:s}) 
is shown by the red dashed lines in Fig. \ref{fig:MHRG}.
\item Include the possibility to describe resonances in the continuum using a Breit-Wigner ansatz and postulate a structureless negative background phase shift which assures the Levinson theorem, namely that at asymptotic large energies the total phase shift in each channel goes to zero. The resonance masses and width as parameters of the Breit-Wigner ansatz are temperature dependent, a simple linear dependence is assumed. 
In this case, 
labelled "SUC" and shown in the bottom panel of Fig. \ref{fig:phase_shift_models},
there is a finite, nonvanishing contribution to the thermodynamics from hadronic resonances in the continuum for temperatures above $T_{\rm Mott}$, see the green solid lines in Fig. \ref{fig:MHRG}.
\end{enumerate}

The simplest model that is in agreement with the Levinson theorem is a SUSD ansatz \cite{Blaschke:2017pvh,Bastian:2018mmc}. It results in the Mott dissociation of hadrons at high temperatures and chemical potentials
\begin{eqnarray}
\label{step-up-step-down}
    \delta_i(M,T,\mu)
    &=&
    \pi
    \Theta(M_{{\rm thr},i}-M_i)\nonumber\\
    &&\left\{
    \Theta(M-M_i)
    -\Theta(M-M_{{\rm thr},i})
    \right\}~.
\end{eqnarray}
Here, the only ingredients are the medium-dependent hadron masses $M_i$ and the sum of the masses of their quark constituents which determine the threshold mass 
$M_{{\rm thr},i}$ of the corresponding continuum of unbound states.
The main drawback of the simple ansatz of Eq.~(\ref{step-up-step-down}) is that it neglects any contribution from the continuum of scattering states.
This ansatz has recently been applied in the description of lattice QCD thermodynamics at $\mu=0$ \cite{Blaschke:2020lhm}. 
In the absence of continuum correlations, the requirement of a smooth transition region between the pure HRG and pure QGP phases for the temperature interval 
$160 \lesssim T[{\rm MeV}] \lesssim 200$ 
was fulfilled by adopting a smooth drop of the quark masses, viz. the continuum threshold over that region of temperatures. This resulted in a sequential dissociation of hadrons with the consequence of a staircase-like fine structure of the thermodynamic potential which entailed a fuzzy behavior of its derivatives. 
Besides this caveat, if the Mott dissociation is related to the chemical freeze-out, this scenario would entail a state-dependent smearing of freeze-out parameters for which there are no indications in heavy-ion collision experiments.

For the present work we take the ansatz of \cite{Wergieluk:2012gd} and extend it from the original $\pi$ and $\sigma$ channels to include contributions from each cluster with the appropriate Mott dissociation effect at high temperatures.
This modified phase shift model reads
\begin{eqnarray}
  &&\delta_i(M)
    =\delta^*_i(M) \left[\Theta(M^2_{{\rm thr},i}-M^2) \right. \nonumber\\
    && \left.+\frac{1}{\pi}{\rm acos}\left(\frac{2 M-2 M_{{\rm thr},i}-N_i\Lambda_i}{N_i\Lambda_i}\right)\right.\nonumber \\&&\left. \times  \Theta(M^2-M^2_{{\rm thr},i}) \Theta((M_{{\rm thr},i}+\Lambda_i N_i)^2-M^2)\right]
\end{eqnarray}
with 
\begin{equation}
\label{eq:phase_shift_extended}
    \delta^*_i(M)
    =
    \begin{cases}
        \pi\, \Theta(M^2-M_i^2)
       ~,&T<T_{\rm Mott}\\
        \pi\frac{\mathrm{atan}\left(\frac{M^2-M_i^2}{\Gamma_iM_i}\right)-\mathrm{atan}\left(\frac{ M_{{\rm thr},i}^2-M_i^2}{\Gamma_iM_i}\right)}{\frac{\pi}{2}-\mathrm{atan}\left(\frac{ M_{{\rm thr},i}^2-M_i^2}{\Gamma_iM_i}\right)}\\ \times
        \Theta(M^2-M_{{\rm thr},i}^2)~.&T>T_{\rm Mott}
    \end{cases}
\end{equation}
For temperatures above $T_{\rm Mott}$, a linear temperature dependence of the mass and the decay width of the resonance is adopted according to Eqs. (\ref{eq:mass}) and (\ref{eq:width}), respectively.

The encoded mass dependence of the phase shift is shown in 
the bottom panel of 
Fig. \ref{fig:phase_shift_models}
for the generic case of the pion at rest in the medium (where $\omega=M$)
for different temperatures around the Mott transition at $T_{\rm Mott}=T_c=156.5$ MeV.

\section{Results for QCD thermodynamics}
\label{sec:results}

\begin{figure}[!t]
    \centering
    \includegraphics[width=\columnwidth]{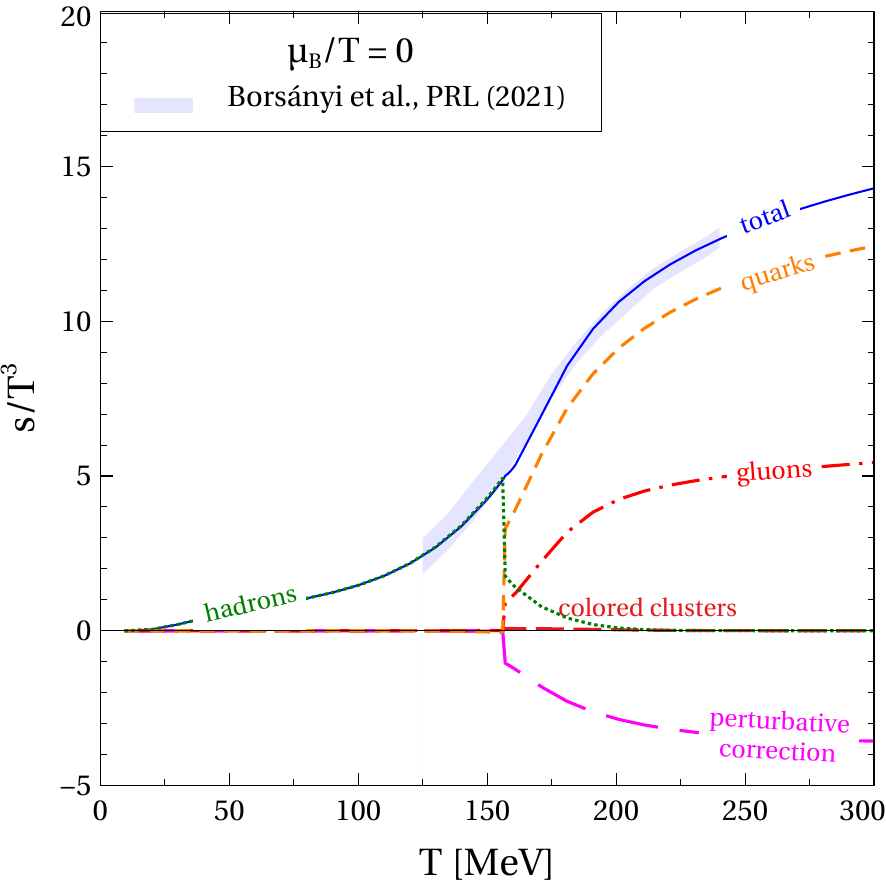}
    \includegraphics[width=\columnwidth]{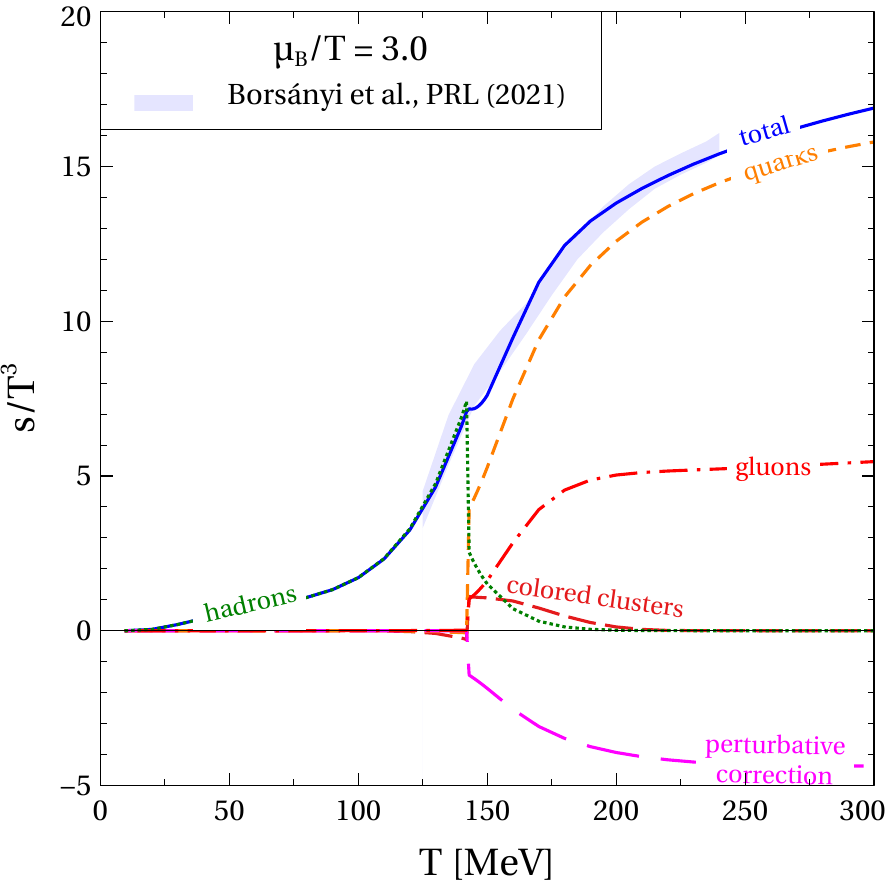}
    \caption{Scaled entropy density $s/T^3$ as a function of temperature $T$ calculated at $\mu_B/T=0$ (upper panel) and $\mu_B/T=3.0$ (lower panel). 
    Partial contributions of hadrons, quarks, gluons as well as perturbative correction and total $s/T^3$ are represented by different curves indicated in figures.
    The shaded regions correspond to the lattice QCD results \cite{Borsanyi:2021sxv} in good agreement with the result of the present model (thick solid line). 
    }
    \label{fig:entropy}
\end{figure}

\begin{figure}[t]
    \centering
    \includegraphics[width=\columnwidth]{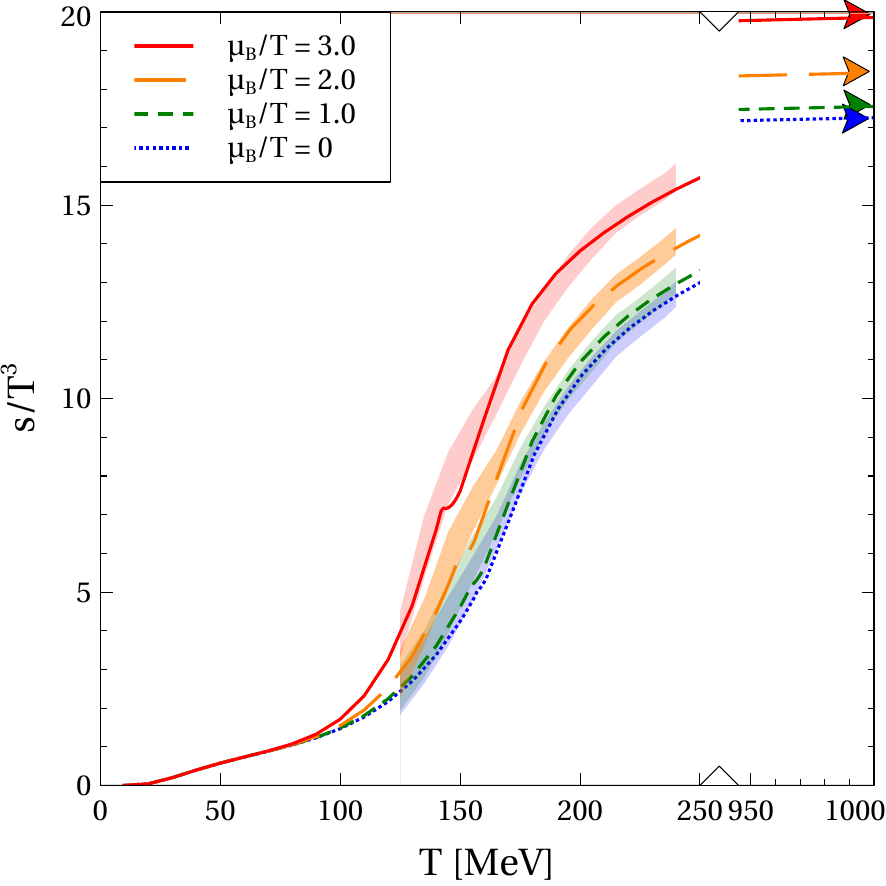}
    \caption{Scaled entropy density $s/T^3$ as a function of temperature $T$ calculated at different values of $\mu_B/T$ indicated in the legend. 
    The shaded regions of the corresponding colors represent the lattice QCD results \cite{Borsanyi:2021sxv}.
    The triangles of corresponding colors are the values obtained within $\mathcal{O}(\alpha_s)$ perturbative QCD, see Tab. \ref{tab:g2}, and are attained for large $T$. Note the broken temperature axis.}
    \label{fig:entropy_all}
\end{figure}

\begin{figure}[!h]
    \centering
    \includegraphics[width=\columnwidth]{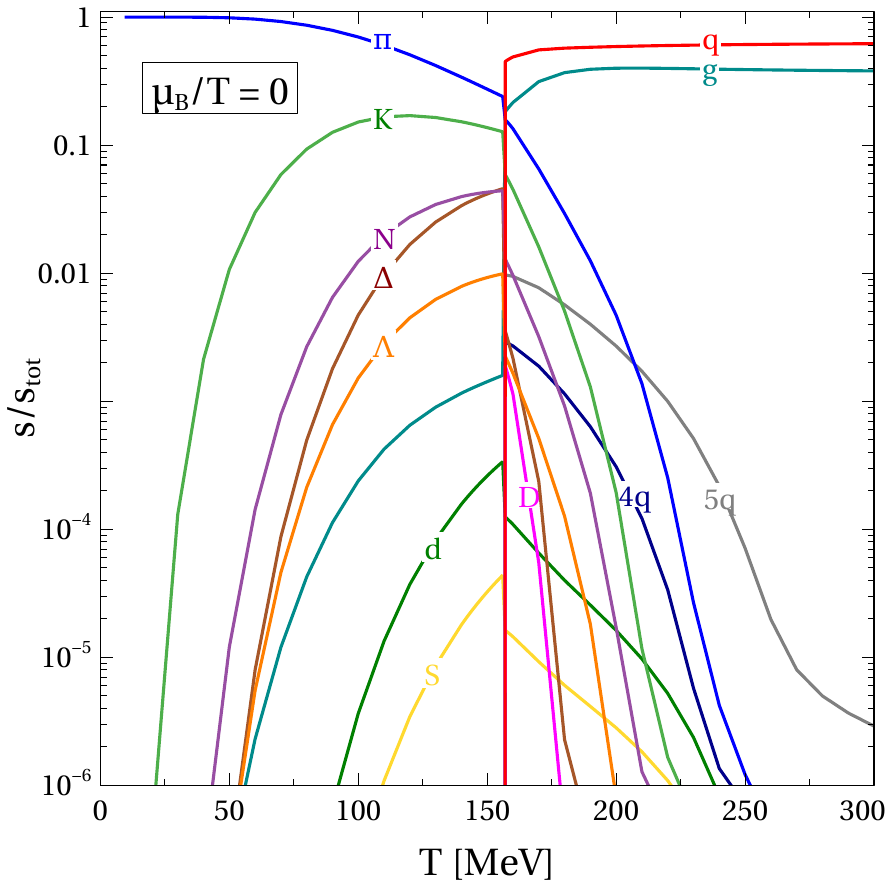}
    \caption{
    Fractions of entropy density carried by different species as a function of temperature $T$ calculated at vanishing baryochemical potential. 
    Notably, the red and turquoise solid lines stands for the entropy fractions of quarks and gluons, resp.,  which are dominant in the QGP phase and strongly suppressed below the Mott transition temperature, because of the effective confinement mechanism in the present work.
    }
    \label{fig:entropy-fractions}
\end{figure}

After having defined the Beth-Uhlenbeck type approach to hadron-quark-gluon matter in Sect. \ref{sec:GBU} and making plausible assumptions about the key ingredients for its numerical evaluation in Sect. \ref{sec:model}, we are now in the position to present results for the entropy density and the net baryon density.
In Fig. \ref{fig:entropy}, we present the scaled entropy $s/T^3$ as function of the temperature. In the upper panel, we show the results for vanishing baryochemical potential, while in the lower panel the case $\mu_B/T=3.0$ is presented and presented to the results of lattice QCD simulations \cite{Borsanyi:2021sxv}, which are given as a grey band.
There is a very good agreement.
On this basis, we can now give an interpretation of the QCD thermodynamics. 
Within the sudden drop model for the quark mass, there is a perfect confinement of quark and gluon degrees of freedom for temperatures below the pseudocritical one, which in our case coincides with the Mott temperature of hadron dissociation $T_{\rm Mott}=T_c=156.5$ MeV.
For $T<T_{\rm Mott}$, the entropy is saturated by the contributions from the ideal hadron resonance gas.
We have  implemented here the same set of hadronic states from the PDG as in Ref. \cite{Albright:2014gva},
and added the deuteron as a hexaquark state as well as the hypothetical sexaquark $S(uuddss)$ which is discussed, e.g., in \cite{Shahrbaf:2022upc} and references therein.
At $T=T_{\rm Mott}$, the hadron contribution drops by about half of its value as a result of the temperature dependence of the hadronic correlations encoded in their phase shift adopted in this model.
The full spectrum of hadronic states contributes to the entropy as continuum correlations modeled by their phase shifts. With increasing temperature, these contributions die out and vanish above $T\sim 200$ MeV.
We note that it is important for obtaining a continuous behaviour of the entropy at the transition temperature that the contributions in the deconfined phase, stemming from quark quasiparticles, gluons, perturbative virial corrections and colored clusters conspire so as to compensate the sudden drop in the hadronic entropy. 
One may attribute this to quark-hadron duality.
A remarkable finding is that at finite chemical potential the colored clusters (diquarks, 4--quark and 5--quark states) play an important role. 

In Fig. \ref{fig:entropy_all}, we demonstrate that the chemical potential dependence of the lattice QCD data is correctly reproduced and that for large temperatures the $\mathcal{O}(\alpha_s)$ perturbative QCD limits are attained.

\begin{figure}[!h]
    \centering
    \includegraphics[width=\columnwidth]{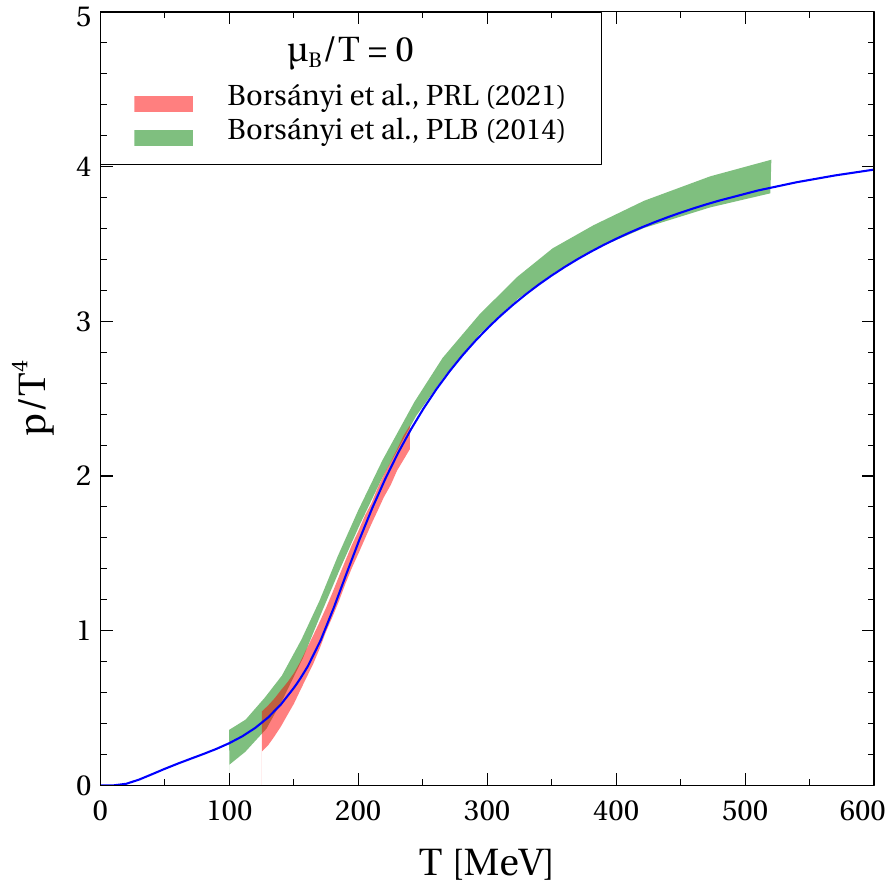}
    \caption{
Scaled pressure $p/T^4$ as a function of temperature $T$ at vanishing baryochemical potential $\mu_B/T=0$ obtained from integrating the entropy (\ref{eq:pressure}) (blue solid line), in comparison with two sets of lattice QCD data from Borsanyi et al. (2014) \cite{Borsanyi:2013bia} and  
Borsanyi et al. (2021) \cite{Borsanyi:2021sxv}, shown as shaded regions.}
    \label{fig:pressure}
\end{figure}

In Fig. \ref{fig:entropy-fractions}, we demonstrate the very power of the present approach. Namely, that we can identify the composition of the matter on both sides of the transition. The fractional contributions of hadron species reflect the composition of the HRG within the statistical model which gives an excellent understanding of hadron production in heavy-ion collision experiments
(see, e.g., \cite{Andronic:2017pug} and references therein). 
We want to underline that the red and turquoise solid lines in Fig. \ref{fig:entropy-fractions} standing for the quark and gluon fractions of the entropy, respectively, shows the degree of realization of quark and gluon confinement in this model. While quarks and gluons dominate the high temperatuure region above 
$T\sim 200$ MeV, their entropy fraction in the hadronic phase below $T_{\rm Mott}$ is below the per cent level. This realization of sudden quark-gluon confinement is due to the combination of the Polyakov-loop suppression of the colored state and the large constituent quark mass in the hadronic phase. 

We want to comment also on the changes that the entropy fractions undergo at the Mott transition.
The contributions of all multiquark clusters get suppressed by about a factor two because of the dramatic change in the generic phase shift function, when they change their character from a bound state to a correlation in the continuum, see Fig. \ref{fig:phase_shift_models}.
This change can be directly read-off from Fig. \ref{fig:entropy-fractions} for color singlet hadrons (pions, kaons, nucleons, ...) while for colored clusters there is the effect of Polyakov loop suppression which acts on their distribution functions in the hadronic phase, where the traced Polyakov loop is close to zero, see Fig. \ref{fig:PL}. 
In the case of diquarks, both effects compensate each other, so that the entropy fraction of diquarks is continuous at the Mott transition, while for the 4--quark and 5--quark states, the Polyakov loop suppression overcompensates the effect of Mott dissociation at  $T_{\rm Mott}$.
The $S(1885)$ contribution is a unique prediction of the present model. We note that in Ref. \cite{Blaschke:2021tul}, predictions have been made for the relative yields of $S(1950)/d$ and $S(1700)/d$ in heavy-ion collisions using a thermal statistical model with excluded volume corrections. Similar to the entropy fraction shown in Fig. \ref{fig:entropy-fractions}, the sexaquark-to-deuteron ratio is of the order one. 

\begin{figure}
    \centering
    \includegraphics[width=\columnwidth]{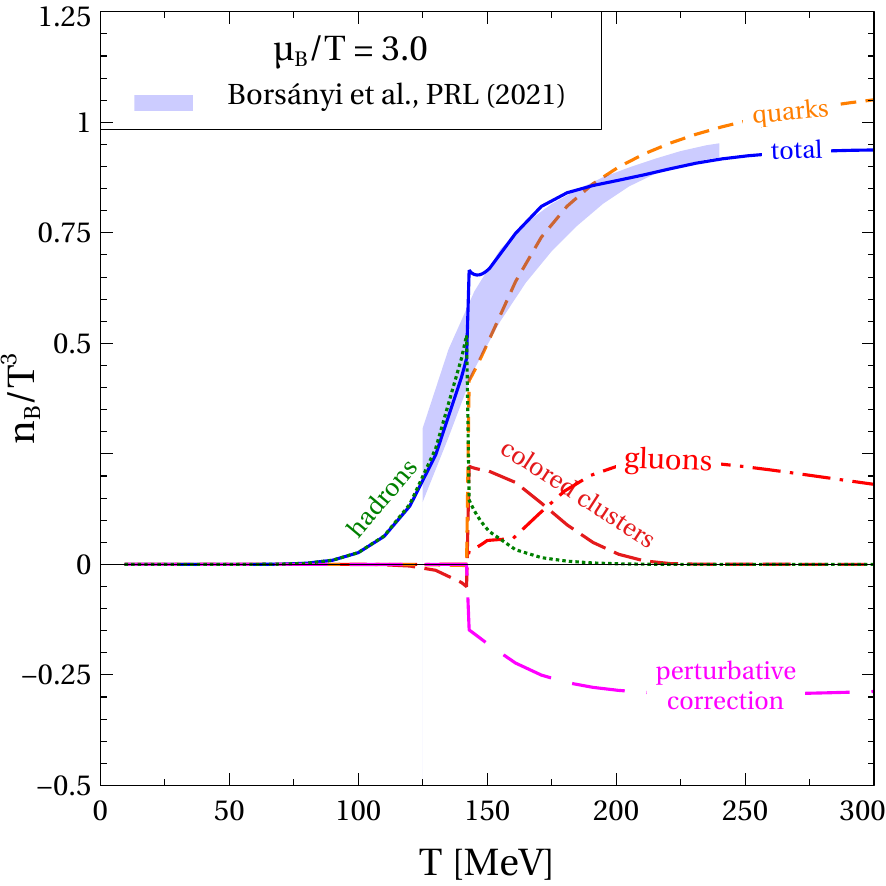}
    \caption{Baryon density as a function of temperature for $\mu_B/T=3.0$ (blue solid line) compared to lattice QCD simulations (light blue band) \cite{Borsanyi:2021sxv}. Shown are also the contributions from partial densities of 
    quarks (orange dashed line), colored clusters (red long dashed line), hadrons (green dotted line) and perturbative QCD
    (magenta very long dashed line).}
    \label{fig:bdens}
\end{figure}

A quantity of central interest for the QCD thermodynamics is the pressure because it plays the role of the thermodynamic potential of the grand canonical ensemble. 
We use Eq. (\ref{eq:pressure}) to obtain the pressure by integrating the entropy density over temperature.
The result is shown in Fig. \ref{fig:pressure}. The comparison to two sets of lattice QCD data by Borsanyi et al. (2014) \cite{Borsanyi:2013bia} and 
Borsanyi et al. (2021) \cite{Borsanyi:2021sxv} shows an excellent agreement.

\begin{figure}
    \centering
    \includegraphics[width=\columnwidth]{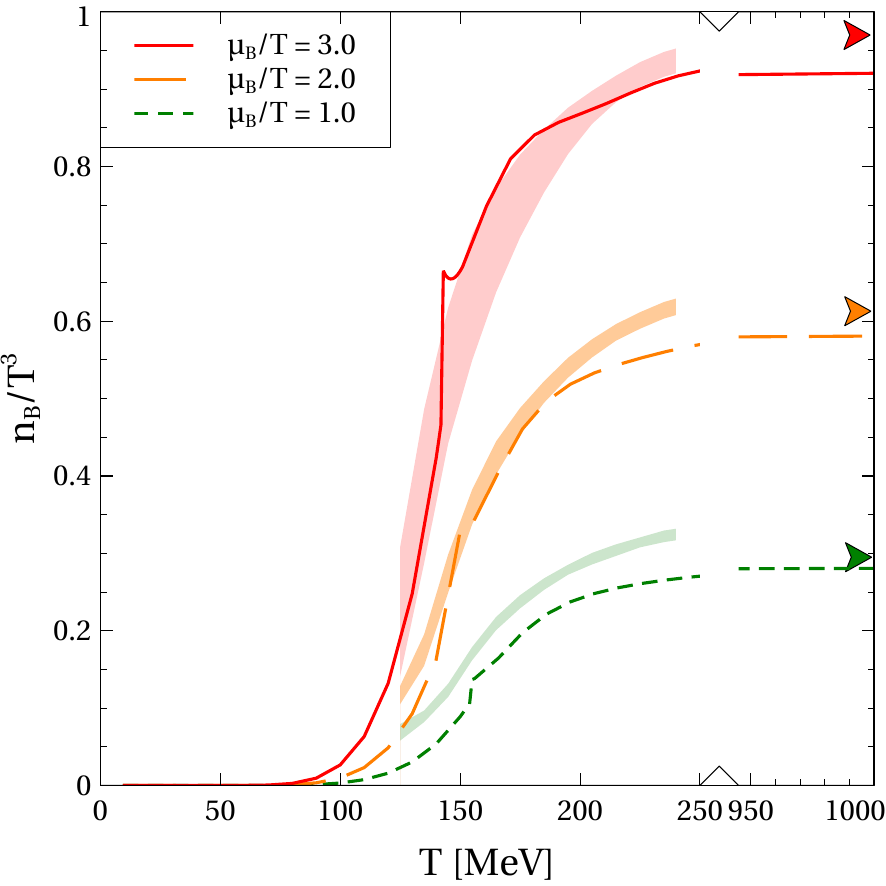}
    \caption{Baryon density as a function of temperature for different values of $\mu_B/T$ compared to lattice QCD simulations \cite{Borsanyi:2021sxv}. 
    The triangles of corresponding colors are the values obtained within $\mathcal{O}(\alpha_s)$ perturbative QCD, see Tab. \ref{tab:g2}, which are attained at large temperature.}
    \label{fig:bdens-3}
\end{figure}

Finally, in Figs. \ref{fig:bdens} and \ref{fig:bdens-3} we show the results for the scaled baryon density as a function of the temperature.
In Fig. \ref{fig:bdens} we show the different components that contribute to the baryon density for $\mu_B/T=3.0$ and compare the resulting total density with the lattice QCD simulation \cite{Borsanyi:2021sxv} shown as a blue band. 
We find it remarkable that the colored clusters contribute more than the color singlet hadrons in the transition region above $T_{\rm Mott}$.
Fig. \ref{fig:bdens-3} demonstrates that the systematics of the chemical potential dependence is correctly accounted for. In particular, the baryon density quickly approaches the $\mathcal{O}(\alpha_s)$ perturbative QCD values while the lattice QCD data slightly overshoot. At temperatures below $T_{\rm Mott}$, where the baryon density is given by the color singlet baryons only, there is a discrepancy with the lattice QCD results which eventually can be related to the known problem of missing strange baryon states in the PDG HRG description \cite{Bazavov:2014xya}. 
Adding baryonic states from a quark model description \cite{Isgur:1978wd} has been demonstrated to solve this problem \cite{Kaczmarek:2022oiu}. 

\section{Summary and Conclusions}
\label{sec:summary}

We have developed a systematic approach to the thermodynamics of the quark-gluon-hadron system that studies quark-quark correlations in the form of bound states and continuum correlations. 
The challenge is to describe, in contrast to conventional plasma systems, the confinement property that allows only color-neutral clusters in the low-density limit. 
This property was taken into account by using the Polyakov-Nambu-Jona-Lasinio model as an effective approach for the many-quark system, 
where the gluon degrees of freedom were eliminated.

We obtained results for arbitrary temperatures and for finite baryonic chemical potentials. Two limiting cases are reproduced: hadronic matter described as hadron resonance gas at low temperatures, for which experimental data are available, and the quark-gluon plasma at high temperatures, for which results from perturbation theory are known. In contrast to semi-empirical approaches such as Ref. \cite{Albright:2014gva}, which use a switch function to interpolate between the two limiting cases, we have outlined a microscopic approach that involves the formation of a condensate, chiral symmetry, and deconfinement. In addition to colored single-quark quasiparticles, colored clusters are also included.

A main topic is the treatment of correlations, for which we employed the correlation functional $\Phi$. Within a cluster decomposition, the special class of "sunset" diagrams was singled out and a generalized Beth-Uhlenbeck equation was used.
This approach allows not only a systematic inclusion of continuum correlations via the scattering phase shifts, but also the inclusion of in-medium effects describing the dissolution of bound states with increasing energy density.

In this work, we have developed the technique of the cluster-virial expansion approach for multi-quark clusters at finite temperatures and chemical potentials based on a cluster generalization of the $\Phi$-derivable approach. 
When a restriction to closed two-loop diagrams in cluster Green's functions is applied to the $\Phi$-functional, this approach is equivalent to the generalized Beth-Uhlenbeck approach to clustering in hot and dense Fermi systems.

The complexity of color confinement in low-density quark matter is accounted for by coupling the quarks and their clusters to the Polyakov-loop background field, which serves to suppress the appearance of colored clusters in the quark confinement region.
We have compared our results for the thermodynamics of clustered quark matter with recent lattice QCD simulations at finite temperature and small chemical potentials where they are currently available and found satisfactory agreement.

Our approach contains some simplifying approximations, especially with respect to the in-medium phase shifts. 
A self-consistent solution of the gap equation in a correlated medium would be of interest to obtain consistent results. 
We expect that such a solution would exhibit a behaviour which justifies the assumption of a sudden switch model for the quark masses that we made in this work.
The few-quark wave equation describing the shift in binding energies and the phase shifts $\delta_i$ is a subject of further investigation. In addition to the sunset diagrams, other diagrams for the correlation functional $\Phi$ can be considered to improve this approximation. Baryon densities and entropies
have been compared with lattice QCD simulations to verify the quality of our approach. Consistent expressions for the various equations of state for the corresponding thermodynamic variables are needed to simulate matter under extreme conditions, such as those encountered in ultrarelativistc heavy-ion collisions and in dense astrophysical objects.

\subsection*{Acknowledgement}
D.B. and M.C. acknowledge support from the Polish National Science Centre (NCN) under grant number\\ 2019/33/B/ST9/03059. 
The work of O.I. was supported by the program Excellence Initiative--Research University of
the University of Wroc?aw of the Ministry of Education and Science.
The work of G.R. received support via a joint stipend from the Alexander von Humboldt Foundation and the Foundation for Polish Science.

\appendix{
\section{Matsubara summations for multiquark Polyakov-loop propagators}
\label{app:1}

\subsection{Diquark propagator and distribution function}
\label{app:1-2}

In this work we restrict ourselves to the consideration of the scalar, color antitriplet diquark. 
The free color antitriplet diquark propagators are obtained after color and Matsubara sum of the product of two color triplet single-quark propagators folded with the antisymmetric Gell-Mann matrices $\lambda_2$, $\lambda_5$ or $\lambda_7$. 
For a better systematics, we will replace these matrices by the completely equivalent totally antisymmetric symbol $\varepsilon_{abc}$ in color space. 
We introduce the bosonic Matsubara frequency $\Omega_m=\omega_n + \omega_n^\prime$ as the sum of the fermionic Matsubara frequencies of the quark propagators that constitute the diquark and obtain the free quark pair propagator.

\begin{eqnarray}
 \label{eq:S_2}
S_2^{(0)} &=& \sum_{ab} \sum_n \varepsilon_{abc} \frac{1}{ (i\omega_n - E_p + \mu)\mathbbm{1}_c - i(A_4)_{aa}} \cdot
\nonumber\\
&&\frac{1}{ (i\omega_{n^\prime} - E_{p^\prime} + \mu)\mathbbm{1}_c - i(A_4)_{bb}}
\nonumber\\
&=& \sum_{ab} \sum_n \frac{\varepsilon_{abc}}{(i\Omega_{m} - E_p - E_{p^\prime} + 2\mu)\mathbbm{1}_c - i(A_4)_{aa} - i(A_4)_{bb}}
\nonumber \\
&&\times \left[ \frac{1}{ (i\omega_n - E_p + \mu)\mathbbm{1}_c - i(A_4)_{aa}} 
\right.
\nonumber\\
&&
\left. +
\frac{1}{ (i(\Omega_m - \omega_{n}) - E_{p^\prime} + \mu)\mathbbm{1}_c - i(A_4)_{bb}} \right]\\
&=& \frac{-\left.f^{(1),+}_\phi\right|_{E_p}+\left.\left[f^{(1),-}_{\phi}\right]^*\right|_{E_{p^\prime}}}{(i\Omega_{m} - E_p - E_{p^\prime} + 2\mu)\mathbbm{1}_c + i(A_4)_{cc}}~,
\end{eqnarray}
where $\mathbbm{1}_c$ is the diagonal matrix in color space and we have employed the result of the summation over color and Matsubara frequency of the single quark propagators given by 
Eq.~(\ref{eq:f-phi}) and used the identity $-(A_4)_{cc} = \varepsilon_{abc} \left[ (A_4)_{aa} + (A_4)_{bb}\right]$.

The diquark distribution function in the presence of the Polyakov-loop background field can now be obtained analogously to the case of the single quark distribution. 
For that we introduce the abbreviation $\omega=E_p + E_{p^\prime}$ for the two-particle energy and neglect the compositeness effect by suppressing the Pauli blocking factor $-\left.f^{(1),+}_\phi\right|_{E_p}+\left.\left[f^{(1),-}_{\phi}\right]^*\right|_{E_{p^\prime}} = 1-\left.f^{(1),+}_\phi\right|_{E_p}-\left.f^{(1),+}_{\phi}\right|_{E_{p^\prime}}\longrightarrow 1$.
By summation over Matsubara frequencies and color index we then obtain
\begin{eqnarray}
f_\phi^{(2),+} &=&  \frac{1}{3} \sum_{c=1}^3  \frac{1}{\beta}\sum_n\frac{1}{i\omega_{n}^{(2)} - \omega \mathbbm{1}_c}\\
&=&  \frac{1}{3} \sum_{c=1}^3 
 \frac{1}{\beta}\frac{\partial \ln \left[1 - y_2^+{\rm e}^{i\beta (A_4)_{cc}}\right]}{\partial \omega}\\
 &=&\frac{1}{3\beta} \frac{\partial}{\partial \omega}  \ln \left\{[1-y_2^+{\rm e}^{-2i\beta \varphi_8} ] [1-y_2^+{\rm e}^{i\beta (\varphi_8-\varphi_3)} ] 
 \right.
 \nonumber\\
&& 
\left.
[1-y_2^+{\rm e}^{i\beta (\varphi_8+\varphi_3)} ] \right\}\\
 &=& \frac{1}{3\beta} \frac{\partial}{\partial \omega}  \ln \left[1 - 3 ({\phi} - \bar{\phi} y_2^+) y_2^+ - {y_2^+}^3\right]~,
 \end{eqnarray}
where $y_2^+(\omega) = \exp[-\beta (\omega-2\mu)]$ and $\omega_{n}^{(2)}=(2n\pi T -i 2\mu)\mathbbm{1}_c + A_4$, see Tab.~\ref{tab:Matsubara}. 
Carrying out the $\omega-$derivative, we obtain the Polyakov-loop generalized diquark distribution function 
 \begin{eqnarray}
 \label{eq:f-phi2}
f_\phi^{(2),+} &=& \frac{({\phi} - 2\bar{\phi} y_2^+) y_2^+ + {y_2^+}^3}{1 - 3 ({\phi} - \bar{\phi} y_2^+) y_2^+ - {y_2^+}^3}~.
\end{eqnarray}
 
\subsection{Three-quark (nucleon) propagator and distribution function}
\label{app:1-3}
We obtain higher order cluster Green's functions as a result of their bipartition into lower lying cluster states which are already known.
In this way, we combine the color antitriplet diquark (anticolor a) with a color triplet quark (color a) in order to obtain a color singlet nucleon, see Tab.~\ref{tab:Matsubara}.
In this way, we obtain the free three-quark (nucleon=quark+diquark) propagator as 
\begin{eqnarray}
 \label{eq:S_3-1}
S_3^{(0)} &=& \sum_{ab} \sum_n \delta_{ab}\frac{1}{ i\omega^{(1)}_n - E_p \mathbbm{1}_c } \cdot
\frac{1}{ i\omega^{(2)}_{n^\prime} - E_{p^\prime}\mathbbm{1}_c }\\
&=& \sum_{a} \sum_n \frac{1}{(i\omega^{(3)}_{m} - (E_p + E_{p^\prime})\mathbbm{1}_c } \left[ \frac{1}{ i\omega^{(1)}_n - E_p\mathbbm{1}_c} 
\right.
\nonumber\\
&&\left. +
\frac{1}{ (i(\omega^{(3)}_m - \omega^{(1)}_{n}) - E_{p^\prime}\mathbbm{1}_c} \right]\\
&=& \frac{-\left.f^{(1),+}_\phi\right|_{E_p}+\left.\left[f^{(2),-}_{\phi}\right]^*\right|_{E_{p^\prime}}}{i\omega^{(3)}_{m} - (E_p + E_{p^\prime})\mathbbm{1}_c}
\nonumber\\
&=& \frac{1-\left.f^{(1),+}_\phi\right|_{E_p}+\left.f^{(2),+}_{\phi}\right|_{E_{p^\prime}}}{i\omega^{(3)}_{m} - (E_p + E_{p^\prime})\mathbbm{1}_c}~.
 \label{eq:S_3-3}
 \end{eqnarray}
Now it is obvious that the nucleon is a color singlet state and no summation over colored degrees of freedom need to be done.
The summation over the fermonic Matsubara frequancy $\omega^{(3)}_{m}=(2m+1)\pi T -i 3 \mu $ is standard and gives
\begin{eqnarray}
f_\phi^{(3)} (\omega) &=& f^{(3)} (\omega) = \frac{1}{\beta} \sum_n\frac{1}{i\omega_{n}^{(3)} - \omega\mathbbm{1}_c}=  \frac{y_3^+}{1+y_3^+}~,
 \end{eqnarray}
where $y_3^+(\omega)=\exp[-\beta (\omega-3\mu)]$ and we have neglected the phase space occupation factor in the numerator
of Eq.~(\ref{eq:S_3-3}).
 
\subsection{Polyakov-loop generalized multi-quark distribution functions}

On the basis of the derivations presented in Appendices \ref{app:1-2} and \ref{app:1-3}, we can  now summarize the generalized multi-quark distribution functions that are obtained for multi-quark states which are dimers of other multiquark states. 
To this end it is sufficient to know the Matsubara frequencies for the $a-$quark cluster and its color structure (triplet, antitriplet or singlet)
as given in Tab.~\ref{tab:Matsubara}.

For the color triplet state with an odd number of net valence quarks, we obtain 
\begin{eqnarray}
 \label{eq:fa-phi-odd}
f^{(a),\pm}_{\phi}&\stackrel{\text{(a odd)}}{=}&\frac{(\bar{\phi} + 2\phi y_a^\pm) y^\pm_a + {y_a^\pm}^3}{1 + 3 (\bar{\phi} + \phi y_a^\pm) y_a^\pm + {y_a^\pm}^3}~, 
\end{eqnarray}
which represents the Polyakov--loop--generalized Fermi function, with notable limiting cases being the single quark distribution function $f^{(1),\pm}_{\phi}$ and the color singlet Fermi function $f^{(a),\pm}_{\phi=1}$.

For an even number of net valence quarks, still in the color triplet state, we obtain 
\begin{eqnarray}
 \label{eq:fa-phi-even}
f^{(a),\pm}_{\phi}&\stackrel{\text{(a even)}}{=}&\frac{(\bar{\phi} - 2{\phi} y_a^\pm) y^\pm_a + {y_a^\pm}^3}{1 - 3 (\bar{\phi} - {\phi} y_a^\pm) y_a^\pm - {y_a^\pm}^3}~,
\\
~y_a^\pm &=& \exp[-\beta (\omega \mp a\mu)] ~,
\end{eqnarray}
which represents the Polyakov--loop--generalized Bose function, with notable limiting cases being the diquark distribution function $\left[f^{(2),\pm}_{\phi}\right]^*$ (the complex conjugation represents a diquark in the color--antitriplet state) and the color singlet Bose function $f^{(a),\pm}_{\phi=1}$.

}
\section{High-temperature $\mathcal{O}(g^2)$ perturbative QCD limit}

An important benchmark for the development of an effective model description that reproduces lattice QCD thermodynamics is the $\mathcal{O}(g^2)$ perturbative QCD limit at high temperatures.
For the readers convenience, we quote here expressions from Appendix B of Ref. \cite{Bollweg:2022fqq}, starting with the pressure at this order, which reads for the three-flavor case ($N_f=3$)
\begin{eqnarray}
    p&=&p_{\rm id} - g^2 \, p_2~,\\
    p_{\rm id}&=& \frac{8\pi^2}{45}T^4 \left(1+\frac{63}{32} \right) + \frac{3}{2}\mu^2 T^2 +\frac{3}{4\pi^2}\mu^4~,\\
    p_2 &=& \frac{1}{6}T^4 \left(1+\frac{5}{4} \right) 
    + \frac{3}{4\pi^2}\mu^2 T^2 +\frac{3}{8\pi^4}\mu^4~.
\end{eqnarray}
Here, we consider the case of vanishing charge and strangeness chemical potentials $\mu_S=\mu_Q=0$, for which $\mu_u=\mu_d=\mu_s=\mu=\mu_B/3$.

From the pressure as the thermodynamic potential, one can derive the other equations of state, such as the density and the entropy density,

\begin{eqnarray}
    n&=& \frac{\partial p}{\partial \mu}\bigg|_{T={\rm const}}\nonumber\\
    &=&3 \mu T^2 + \frac{3}{\pi^2} \mu^3 
    - g^2 \left( \frac{3}{2\pi^2} \mu T^2 + \frac{3}{2\pi^4} \mu^3 \right),\\
    s&=&\frac{\partial p}{\partial T}\bigg|_{T={\rm const}} \nonumber\\
    &=& \frac{32\pi^2}{45}T^3 \left(1+\frac{63}{32} \right) + 3\mu^2 T \nonumber\\
    && - g^2 \left[\frac{2}{3}T^3 \left(1+\frac{5}{4} \right) 
    + \frac{3}{2\pi^2}\mu^2 T \right]~.
\end{eqnarray}
In table \ref{tab:g2}, we give the limiting values for the dimensionless baryon density $n_B/T^3$ and entropy
$s/T^3$ to $\mathcal{O}(g^2)$ for four values of $\hat{\mu}_B=\mu_B/T=0$, 1.0, 2.0 and 3.0, for which there are results from lattice QCD simulations \cite{Borsanyi:2021sxv} available.

\begin{table}[h]
    \centering
    \begin{tabular}{|c|c|c|}
\hline
              & $n_B/T^3=$ & $s/T^3=\frac{19\pi^2}{9}\left(1-\frac{27g^2}{38\pi^2} \right)$ \\
      $\hat{\mu}_B$  & $\left(\frac{\hat{\mu}_B}{3} +\frac{\hat{\mu}_B^3}{27\pi^2}\right)\left(1-\frac{g^2}{2\pi^2} \right)$&  
      $+\frac{\hat{\mu}_B^2}{3}\left(1-\frac{g^2}{2\pi^2} \right)$
      \\
      \hline
        0.0 & 0.0   &  17.3072 \\  
        1.0 & 0.2969&  17.6008 \\
        2.0 & 0.6137&  18.4816 \\
        3.0 & 0.9701&  19.9497 \\
\hline
          \end{tabular}
    \caption{limiting values for the dimensionless baryon density $n_B/T^3$ and entropy
$s/T^3$ to $\mathcal{O}(g^2)$ in perturbative QCD for 
$\alpha_s=g^2/(4\pi)=0.1872$.}
    \label{tab:g2}
\end{table}
These values are approached asymptotically by our model and by the lattice QCD simulations, as can be seen in Figs. 
\ref{fig:entropy_all} and \ref{fig:bdens-3}, where they are indicated by small right arrows of the corresponding colors.


\end{document}